\documentclass[11pt]{article}

\usepackage{epsf}
\usepackage{verbatim}
\usepackage{makeidx}
\usepackage{graphicx}
\usepackage{amssymb}
\usepackage{amsmath}
\usepackage{latexsym}
\usepackage{amsbsy}
\usepackage{epsfig}

\usepackage{chicagob}

\usepackage{endnotes}

\usepackage{theorem}

\newtheorem{theorem}{Theorem}
\newtheorem{definition}[theorem]{Definition}
\newtheorem{example}[theorem]{Example}

\newcommand{\commentout}[1]{}

\begin{document}
\bibliographystyle{chicago}

\title{Algorithmic Information Theory}
\author{Peter D. Gr\"unwald \\ CWI, P.O. Box 94079 \\
NL-1090 GB Amsterdam, The Netherlands \\
E-mail: pdg@cwi.nl  
\and Paul M.B. Vit\'anyi \\
CWI , P.O. Box 94079\\
 NL-1090 GB Amsterdam \\
The Netherlands \\
E-mail: pdg@cwi.nl
}

\maketitle
\begin{abstract}
  We introduce {\em algorithmic information theory}, also known as the
  theory of {\em Kolmogorov complexity}.  We explain the
  main concepts of this quantitative approach to defining
  `information'.  We discuss the extent to which Kolmogorov's and
  Shannon's information theory have a common purpose, and where they
  are fundamentally different. We indicate how recent developments
  within the theory allow one to formally distinguish between {\em
    `structural' (meaningful) and `random' information\/} as measured
  by the {\em Kolmogorov structure function}, which leads to a
  mathematical formalization of Occam's razor in inductive inference.
  We end by discussing some of the philosophical implications of the
  theory.
\end{abstract}
\paragraph{Keywords} 
Kolmogorov complexity, algorithmic information theory, Shannon
  information theory, mutual information, data compression, Kolmogorov structure function, Minimum Description Length Principle.

\newlength{\ownboxwidth} \newlength{\ownleftboxwidth}
  \newlength{\ownrightboxwidth} \setlength{\ownboxwidth}{\textwidth}
  \addtolength{\ownboxwidth}{-2.5 cm} \setlength{\ownleftboxwidth}{0.1
    cm} \setlength{\ownrightboxwidth}{0.01 cm}
 
\newcommand{\prs}{\par}
\newcommand{\emptyline}{\vspace{\baselineskip}}
\newcommand{\halfemptyline}{\vspace{.25cm}}

\newcounter{repeatproba}
\newcounter{repeatprobb}
\newcounter{repeatprobc}
\newcounter{lemmastronglawcounter}
\newcounter{snavel}

\newcounter{myenumeratecounter}
\newenvironment{myenumerate}{\begin{list}
{\arabic{myenumeratecounter}.}
{
\usecounter{myenumeratecounter}
\setlength{\itemsep}{0.0 cm}
}
}
{\end{list}}

\newcommand{\ownbox}[2]{
\noindent \ \vspace{\baselineskip} \\
\noindent
\fbox{\hspace*{\ownleftboxwidth} \parbox{\ownboxwidth}{\
\\ %\normalsize
{\small \bf #1
} \\ \noindent \small #2 }\hspace*{\ownrightboxwidth}
} \ \vspace{\baselineskip} \\}

\newcommand{\myargmax}[1]{\ensuremath{\underset{#1}{\ \arg \max \ }}}
\newcommand{\myargmin}[1]{\ensuremath{\underset{#1}{\ \arg \min \ }}}
\newcommand{\isbydefinition}{\ensuremath{{:=}}}
%pdg: \isbydefinition changed 15-8-03. Was '\equiv'
\newcommand{\stsum}{\sum_{i=1}^n}
\newcommand{\transpose}{\ensuremath{{\text{\sc T}}}}
\newcommand{\indicator}{\ensuremath{{\mathbf 1}}}

\newcommand{\distance}{\ensuremath{{d}}}

\newcommand{\freq}{\ensuremath \gamma}
\newcommand{\remainder}{\ensuremath R}
\newcommand{\constant}{\ensuremath{K}}
\newcommand{\const}{\ensuremath{c}}

% SETS AND THEIR ELEMENTS 

% SETS 1: general sets

\newcommand{\reals}{{\mathbb R}}
\newcommand{\booleans}{{\mathbb B}}
\newcommand{\integers}{{\mathbb Z}}
\newcommand{\naturals}{{\mathbb N}}
\newcommand{\simplexN}[1]{\ensuremath{\Delta}^{(#1)}}

% TOPOLOGY

\newcommand{\closure}{\ensuremath{\mathrm{cl \ }}}
\newcommand{\interior}{\ensuremath{\mathrm{int \ }}}
\newcommand{\ball}{\ensuremath{B}}

% SETS 2: Outcomes and Sample Spaces.

\newcommand{\justaset}{\ensuremath {\mathcal U}}
\newcommand{\justanevent}{\ensuremath {\mathcal E}}
\newcommand{\justasymbol}{\ensuremath {s}}
% \justasymbol denoted to use an element of sample space in contexts
% where x is already used (ie Does s occur in sample x_1, ..., x_n ?)

% we use l for dimensionality of sample space, k for dimensionality of
% parameter space, and m for dimensionality of general space, if we do not
% want to commit to an interpretation (eg beginning of chapter 2)

\newcommand{\data}{\ensuremath D}
\newcommand{\Dtrain}{\ensuremath D_{\text{train}}}
\newcommand{\Dtest}{\ensuremath D_{\text{test}}}
%needed for treatment of cross-validation

\newcommand{\dataalphabet}{\ensuremath \mathcal A}
\newcommand{\samplespace}{\ensuremath {{\mathcal X}}}
%peter 22-7-03: note Xspace and samplespace now the same symbol!
\newcommand{\xspace}{\ensuremath {{\mathcal X}}}
\newcommand{\yspace}{\ensuremath {{\mathcal Y}}}
\newcommand{\zspace}{\ensuremath {{\mathcal Z}}}
\newcommand{\outcomex}{\ensuremath x}
\newcommand{\outcomey}{\ensuremath y}
\newcommand{\bogus}{\ensuremath \Box}
\newcommand{\maxoutcome}{\ensuremath m}
\newcommand{\region}{\ensuremath{{\mathbf R}}}

%SETS AND ELEMENTS 3: estimators/hypotheses/parameters

%\newcommand{\mldim}[1]{\ensuremath{\hat{\theta}_{#1}}}
\newcommand{\mlN}[1]{\ensuremath{\hat{\theta}^{(#1)}}}
\newcommand{\ml}{\ensuremath \hat{\theta}}
\newcommand{\mapest}{\ensuremath \breve{\theta}}
\newcommand{\param}{\ensuremath {\theta}}
\newcommand{\paramN}[1]{\ensuremath {\theta}^{(#1)}}
\newcommand{\meanest}{\ensuremath \bar{\theta}}
\newcommand{\truepar}{\ensuremath \theta^*}
\newcommand{\trueprob}{\ensuremath \prob^*}

\newcommand{\twop}[1]{\ensuremath{{\ddot{#1}}}} 
\newcommand{\twopN}[2]{\ensuremath{{\ddot{#1}^{(#2)}}}}
\newcommand{\mcparam}[1]{\ensuremath{\theta_{[1|#1]}}}
\newcommand{\mcmlparam}[1]{\ensuremath{\hat{\theta}_{[1|#1]}}}
\newcommand{\mctrueparam}[1]{\ensuremath{{\theta}^*_{[1|#1]}}}
\newcommand{\mcmlparamb}[2]{\ensuremath{\hat{\theta}_{[#1|#2]}}}
\newcommand{\utm}{\ensuremath{\text{\sc{ul}}}}

\newcommand{\phyp}{\ensuremath {H}}
\newcommand{\ghyp}{\ensuremath {\mathcal H}}
%phyp = point hypothesis
%ghyp = general (point or compound) hypothesis
\newcommand{\hfun}{\ensuremath {h}}

%SETS AND ELEMENTS 4: parameter SETS

%\newcommand{\parasetN}[1]{\ensuremath{\Theta_{#1}}}
\newcommand{\parasetN}[1]{\ensuremath{\Theta^{(#1)}}}
\newcommand{\paraset}{\ensuremath{\Theta}}
\newcommand{\twopparaset}[2]{\ensuremath{{\ddot{\Theta}^{(#1)}_{#2}}}}

% codes and codelengths

\newcommand{\clengthnr}{\ensuremath{l}}

\newcommand{\lunif}{\ensuremath{L_{U}}}
\newcommand{\punif}{\ensuremath{P_{U}}}
\newcommand{\Lint}{\ensuremath{L_{\naturals}}}
\newcommand{\punivint}{\ensuremath{\bar{P}_{\naturals}}}
\newcommand{\indexcode}{\ensuremath C_{\text{index}}}
\newcommand{\shannoncode}{\ensuremath C_{\text{Shannon}}}
\newcommand{\Lindex}{\ensuremath L_{\text{index}}}
\newcommand{\Lshannon}{\ensuremath L_{\text{Shannon}}}
\newcommand{\prefixlengths}{\ensuremath{\mathcal{L}}}
\newcommand{\code}{\ensuremath C}
\newcommand{\CN}[1]{\ensuremath C^{(#1)}}
\newcommand{\codelength}{\ensuremath{L}}
\newcommand{\codelengths}{\ensuremath{\mathcal{L}}}
\newcommand{\regret}{\ensuremath{\mathcal R}}
\newcommand{\maxregret}{\ensuremath{\mathcal R}_{\max}}

% functions, vectors, matrices

\newcommand{\justafunction}{\ensuremath{g}}
\newcommand{\justavector}{\ensuremath{v}}
\newcommand{\justamatrix}{\ensuremath{M}}
% used when we do not want to enforce the interpretation of a parameter space etc.

% probability things

\newcommand{\empprob}{\ensuremath{\mathbb P}}

\newcommand{\prob}{\ensuremath P}
\newcommand{\pd}{\ensuremath P}
\newcommand{\densityf}{\ensuremath{f}}

\newcommand{\Exp}{\ensuremath{\text{\rm E}}}
\newcommand{\cov}{\ensuremath{\text{\rm cov}}}
\newcommand{\var}{\ensuremath{\text{\rm var}}}
\newcommand{\noiseterm}{\ensuremath{Z}}

\newcommand{\expprior}{\ensuremath{q}}

%models 

\newcommand{\Bernoulli}{\ensuremath{\mathcal{B}}}
\newcommand{\BernoulliN}[1]{\ensuremath{{\mathcal{B}}^{(#1)}}}
\newcommand{\fakemarkov}{\ensuremath{\dddot{\Bernoulli}}}
\newcommand{\M}{\ensuremath {{\cal M}}}
\newcommand{\cH}{\ensuremath {{\cal H}}}
\newcommand{\MN}[1]{\ensuremath {{\cal M}^{(#1)}}}
\newcommand{\HN}[1]{\ensuremath {{\cal H}^{(#1)}}}
\newcommand{\cM}{\ensuremath {{\cal C}}}
  %these are 'candidate hypotheses'

% Fisher inf stuff

\newcommand{\observedmatrix}{\ensuremath \hat{I}}
%H already used for entropy
\newcommand{\fishermatrix}{\ensuremath I}
\newcommand{\curvelength}[1]{\ensuremath{\int_{#1} \fisherdens d\theta }}
\newcommand{\fisherdens}{\ensuremath{\sqrt{| \fishermatrix(\theta)|}}}
\newcommand{\fishervolume}{\ensuremath{V_{\text{d}}}}

% CHAPTER 3 

\newcommand{\level}{\mbox{\sc level}}
\newcommand{\descendantset}{\mbox{\sc d}}
\newcommand{\precision}{\ensuremath{d}}

\newcommand{\kl}{\mbox{$D$}}
\newcommand{\entropy}{{\ensuremath{\text{\rm H}}}}

% CHAPTER 4 

% CHAPTER 6, 7

\newcommand{\jp}{\ensuremath{w_{\text{Jef}}}}
\newcommand{\Lref}{\ensuremath{\mathcal{L}_{\text{ref}}}}
\newcommand{\pmodel}{\ensuremath{\theta}}
\newcommand{\prior}{\ensuremath{w}}
\newcommand{\discreteprior}{\ensuremath{W}}
\newcommand{\decision}{\ensuremath{\delta}}
\newcommand{\decisionspace}{\ensuremath{\Delta}}
\newcommand{\loss}{\ensuremath{\text{\sc loss}}}
\newcommand{\emptysequence}{\ensuremath{\text{\sc doen}}}
\newcommand{\realprob}{\ensuremath{\text{\sc doen}}}
\newcommand{\popt}{\ensuremath{\text{\sc doen}}}
\newcommand{\opthypothesis}{\ensuremath{\text{\sc doen}}}
%MOET ANDERS!

\newcommand{\puniv}{\ensuremath{\bar{P}}}
\newcommand{\punivtp}{\ensuremath{\bar{P}_{\text{2-p}}}}
\newcommand{\punivplugin}{\ensuremath{\bar{P}_{\text{plug-in}}}}
\newcommand{\punivcrude}{\ensuremath{\bar{P}_{\text{crude}}}}
\newcommand{\punivref}{\ensuremath{\bar{P}_{\text{refined}}}}
\newcommand{\punivseq}{\ensuremath{\bar{P}_{\text{\rm seq}}}}
\newcommand{\punivseqn}[1]{\ensuremath{\bar{P}_{\text{seq},#1}}}
\newcommand{\punivopt}{\ensuremath{\bar{P}_{\text{\rm opt}}}}
\newcommand{\optprior}{\ensuremath{{\prior}_{\text{\rm opt}}}}
\newcommand{\punivrnml}{\ensuremath{\bar{P}_{\text{\rm rnml}}}}
\newcommand{\punivnml}{\ensuremath{\bar{P}_{\text{\rm nml}}}}
\newcommand{\punivmeta}{\ensuremath{\bar{P}_{\text{\rm meta}}}}
\newcommand{\punivpreq}{\ensuremath{\bar{P}_{\text{\rm preq}}}}
\newcommand{\punivnmln}[1]{\ensuremath{\bar{P}_{\text{\rm nml},#1}}}
\newcommand{\punivbayes}{\ensuremath{\bar{P}_{\text{\rm Bayes}}}}
\newcommand{\luniv}{\ensuremath{\bar{L}}}
\newcommand{\lunivtp}{\ensuremath{\bar{L}_{\text{\rm 2-p}}}}
\newcommand{\lunivnml}{\ensuremath{\bar{L}_{\text{\rm nml}}}}
\newcommand{\lunivbayes}{\ensuremath{\bar{L}_{\text{\rm Bayes}}}}

\newcommand{\volume}{\ensuremath{V_{\text{p}}}}
\newcommand{\Mdisc}{\ensuremath{\ddot{\mathcal M}}}
\newcommand{\probset}{\ensuremath{{\mathcal P}}}
\newcommand{\complexity}{\ensuremath{\text{\bf COMP}}}
\newcommand{\rcomplexity}{\ensuremath{\text{\bf RCOMP}}}
\newcommand{\typset}{\ensuremath{{\mathcal A}}}

%----------------------------------------------------------
% POST chapter 7

%MML

\newcommand{\wftwopartlength}{\ensuremath{L_{\ensuremath{\text{mml-wf}}}}}
\newcommand{\mmlplength}{\ensuremath{L_{\ensuremath{\text{mml-p}}}}}

%MaxEnt
\newcommand{\constraint}{\ensuremath{\mathcal C}}
\newcommand{\constraintdataset}{\ensuremath {{\mathbf C}^n}}
\newcommand{\entropified}[1]{\ensuremath{{\langle {#1} \rangle}}}
\newcommand{\function}{\ensuremath \phi}
\newcommand{\range}{\ensuremath \mathbf U}
\newcommand{\expval}{\ensuremath t}
\newcommand{\mep}{\ensuremath{\prob_{me}}}
\newcommand{\mept}{\ensuremath{\prob_{me,\phi}(\cdot | t)}}
\newcommand{\freqdataset}{\ensuremath {{\mathbf G}^n}}
\newcommand{\freqdatasetn}[1]{\ensuremath {{\mathbf G}^{#1}}}
\newcommand{\constraintprobset}{\ensuremath {{\mathbf M}}}
\newcommand{\constraintdatasetn}[1]{\ensuremath {{\mathbf C}^{#1}}}
\newcommand{\empiricalconstraint}{\ensuremath {{\cal C}_{ e}}}
\newcommand{\average}[2]{\ensuremath \overline{#1}^{#2}}

% Unclear wether needed

%\newcommand{\maxnumclass}{\ensuremath K}
%\newcommand{\numcomps}{\ensuremath m}
%\newcommand{\numdimensions}{\ensuremath k}
%\newcommand{\traindata}{\ensuremath{D}}
\newcommand{\lagrange}{\ensuremath \beta}

% priors

%\newcommand{\jprior}{\ensuremath{\pi}}

%old

%\index{symbols}{zzz@$\tilde{c}$ where $c \in \{
%\pmodel,\lagrange,\hypothesis,\sigma \}$: model that minimizes expected error}
%\index{symbols}{zzz@$\hat{c}$ where $c \in \{
%\pmodel,\lagrange,\hypothesis,\sigma \}$: model that minimizes
%empirical error/maximizes likelihood}
%\index{symbols}{zzz@${c}_{mdl}$ where $c \in \{
%\pmodel,\lagrange,\hypothesis,\sigma \}$: model that minimizes
%two-part codelength}
%\index{symbols}{zzz@$\ddot{c}$ where $c \in \{
%\pmodel,\lagrange,\hypothesis \}$: (unspecified) estimator}
%\index{symbols}{zzz@$\breve{c}$ where $c \in \{
%\pmodel,\lagrange,\hypothesis \}$: Bayesian MAP estimator}
%\index{symbols}{zzz@$\bar{c}$ where $c \in \{
%\pmodel,\lagrange,\hypothesis \}$: mean of Bayesian posterior}
%\index{symbols}{zzz@$c^*$ where $c \in \{
%\pmodel,\lagrange,\hypothesis \}$: `true' model generating the data}

%\newcommand{\convergesatrate}{\ensuremath{\lesssim}}

\newcommand{\lea}{\stackrel{{}_+}{<}}
\newcommand{\gea}{\stackrel{{}_+}{>}}
\newcommand{\eqa}{\stackrel{{}_+}{=}}
\newcommand{\soph}{\mbox{\rm soph}}
\newcommand{\dR}{\ensuremath{\mathbb R}}

\section{Introduction}
How should we measure the amount of information about a phenomenon
that is given to us by an observation concerning the phenomenon?  Both
`classical' (Shannon) information theory (see the chapter by \citeN{HarremoesT06})
and algorithmic information theory start with the idea that this
amount can be measured by {\em the minimum number of bits needed to
  describe the observation}.  But whereas Shannon's 
theory considers description methods that are optimal relative to
some given probability distribution, Kolmogorov's algorithmic
theory takes a different, nonprobabilistic approach: any
computer program that first computes (prints) the string representing
the observation, and then terminates, is viewed as a valid
description. The amount of information in the string is then defined
as the size (measured in bits) of the {\em shortest\/} computer
program that outputs the string and then
terminates. A similar definition can be given for infinite strings,
but in this case the program produces element after element forever.
Thus, a long sequence of 1's such as
\begin{equation}
\label{eq:verkouden}
 \overbrace{11 \ldots 1}^{10000 \mbox{{\scriptsize  \ times}}}
\end{equation}
contains little information because a program of size about $\log 10000$ bits outputs it:
$$
{\tt for} \ i \  {\tt := } \ 1 \ {\tt to} \ 10000 \ ; \ {\tt print} \  1.
$$
Likewise, the transcendental number $\pi = 3.1415...$, an infinite
sequence of seemingly `random' decimal digits, contains but a few bits
of information (There is a short program that produces the consecutive
digits of $\pi$ forever). 
%On the other hand, it will be seen that with overwhelming probability, a sequence of 10000 `truly' random bits, generated by independent tosses of a fair coin, cannot be substantially compressed: the shortest program that prints it will look something like:
%and have length comparable to the length of the string itself. 

Such a definition would appear to make the amount of information in a
string (or other object) depend on the particular programming language
used.  Fortunately, it can be shown that all reasonable choices of
programming languages lead to quantification of the amount of
`absolute' information in individual objects that is invariant up to
an additive constant. We call this quantity the `Kolmogorov
complexity' of the object. While regular strings have small Kolmogorov
complexity, random strings have Kolmogorov complexity about equal to
their own length. Measuring complexity and information in terms of
program size has turned out to be a very powerful idea with
applications in areas such as theoretical computer science, logic,
probability theory, statistics and physics.
\paragraph{This Chapter}
Kolmogorov complexity was introduced independently and with different
motivations by R.J. Solomonoff (born 1926), A.N.  Kolmogorov
(1903--1987) and G. Chaitin (born 1943) in 1960/1964, 1965 and 1966
respectively \cite{Solomonoff64,Kolmogorov65,Chaitin66}. During the
last forty years, the subject has developed into a major and mature
area of research. Here, we give a brief overview of the subject geared
towards an audience specifically interested in the philosophy of
information.  With the exception of the recent work on the Kolmogorov
structure function and parts of the discussion on philosophical
implications, all material we discuss here can also be found in the
standard textbook \cite{LiV97}. The chapter is structured as follows:
we start with an introductory section in which we define Kolmogorov
complexity and list its most important properties. We do this in a
much simplified (yet formally correct) manner, avoiding both
technicalities and all questions of motivation (why this definition
and not another one?).  This is followed by Section~\ref{sec:overview}
which provides an informal overview of the more technical topics
discussed later in this chapter, in
Sections~\ref{sec:kolmogorov}--~\ref{sec:meaning}.  The final
Section~\ref{sec:philosophy}, which discusses the theory's
philosophical implications, as well as Section~\ref{sec:mdl}, which
discusses the connection to inductive inference, are less technical
again, and should perhaps be glossed over before delving into the
technicalities of Sections~\ref{sec:kolmogorov}--~\ref{sec:meaning}.
\section{Kolmogorov Complexity: Essentials}
\label{sec:essentials}
The aim of this section is to introduce  our main notion in the
fastest and simplest possible manner, avoiding, to the extent that
this is possible, all technical and motivational issues. 
Section~\ref{sec:koldef} provides a simple definition of Kolmogorov
complexity. We list some of its key properties in
Section~\ref{sec:key}. Knowledge of these key properties is an
essential prerequisite for understanding the advanced topics treated
in later sections.
\subsection{Definition} 
\label{sec:koldef}
The Kolmogorov complexity $K$ will be defined as a function from
finite binary strings of arbitrary length to the natural numbers
$\naturals$. Thus, $K: \{0,1\}^* \rightarrow \naturals$ is a function
defined on `objects' represented by binary strings. Later the
definition will be extended to other types of objects such as numbers
(Example~\ref{ex:uncomputable}), sets, functions and probability
distributions (Example~\ref{ex:kcgeneralobjects}).
 
As a first approximation, $K(x)$ may be thought of as the length of
the shortest computer program that prints $x$ and then halts.  This
computer program may be written in Fortran, Java, LISP or any other {\em
  universal programming language}. By this we mean a general-purpose programming language in which a universal Turing Machine can be implemented. Most languages encountered in practice have this property. For concreteness, let us fix some universal
language (say, LISP) and define Kolmogorov complexity with respect to it.
The {\em invariance theorem\/} discussed below implies that it does
not really matter which one we pick.  Computer programs often make use
of data. Such data are sometimes listed inside the program. An example
is the bitstring {\tt "010110..."} in the program
\begin{equation}
{\tt print "01011010101000110...010"}
\end{equation}
In other cases, such data are given as additional input to the
program. To prepare for later extensions such as conditional
Kolmogorov complexity, we should allow for this possibility as well.
We thus extend our initial definition of Kolmogorov complexity by
considering computer programs with a very simple input-output
interface: programs are provided a stream of bits, which, while
running, they can read one bit at a time. There are no end-markers in
the bit stream, so that, if a program $p$ halts on input $y$ and
outputs $x$, then it will also halt on any input $yz$, where $z$ is a
continuation of $y$, and still output $x$.  We write $p(y) = x$ if, on
input $y$, $p$ prints $x$ and then halts.  
We define the
Kolmogorov complexity relative to a given language as the length of
the shortest program $p$ {\em plus input\/} $y$, such that, when given
input $y$, $p$ computes (outputs) $x$ and then halts. Thus:
\begin{equation}
\label{eq:kolmogorova}
K(x) := \min_{y,p: p(y) = x} l(p) + l(y),
\end{equation}
where $l(p)$ denotes the length of input $p$, and $l(y)$ denotes the
length of program $y$, both expressed in bits. To make this definition
formally entirely correct, we need to assume that the program $P$ runs
on a computer with unlimited memory, and that the language in use has
access to all this memory. Thus, while the definition
(\ref{eq:kolmogorova}) can be made formally correct, it does obscure
some technical details which need not concern us now. We return to
these in Section~\ref{sec:kolmogorov}.
\subsection{Key Properties of Kolmogorov Complexity}
\label{sec:key}
To gain further intuition about $K(x)$, we now list five of its key
properties. Three of these concern the size of $K(x)$ for commonly
encountered types of strings. The fourth is the invariance theorem,
and the fifth is the fact that $K(x)$ is uncomputable in general. Henceforth, we use $x$ to denote finite bitstrings. We
abbreviate $l(x)$, the length of a given bitstring $x$, to $n$. We 
use boldface ${\bf x}$ to denote an infinite binary string.
In that case, $x_{[1:n]}$ is used to denote the initial $n$-bit
segment of ${\bf x}$.

\paragraph{1(a). Very Simple Objects: $K(x) = O(\log n)$.}
$K(x)$ must be small for `simple' or `regular' objects $x$. For
example, there exists a fixed-size program that, when input $n$,
outputs the first $n$ bits of $ \pi$ and then halts. As is easy to see
(Section~\ref{sec:codingprelim}), specification of $n$ takes $O(\log n)$
bits.
%To see this, note that
%to specify a decimal number between $10^k$ and $10^{k-1}$, we need $k$
%digits. Similarly, to specify a binary number between $2^k$ and
%$2^{k-1}$, we need $k$ bits. Taking logarithms to base 2, it follows
%we need approximately $\log n$ bits to specify $n$.  This reasoning is
%not entirely correct, for reasons explained in Section~\ref{sec:coding}:
%in reality we  need $\log n + O( slightly more bits, but still $O(\log n)$. 
Thus, when $x$ consists of the first $n$ bits of $\pi$, its complexity
is $O(\log n)$.  Similarly, we have $K(x) = O(\log n)$ if $x$
represents the first $n$ bits of a sequence like (\ref{eq:verkouden})
consisting of only 1s. We also have $K(x) = O (\log n)$ for the first
$n$ bits of $e$, written in binary; or even for the first $n$ bits of
a sequence whose $i$-th bit is the $i$-th bit of $e^{2.3}$ if the
$i-1$-st bit was a one, and the $i$-th bit of $1/\pi$ if the $i-1$-st
bit was a zero. 
\commentout{BOEK
 A string can be very
`complex' in the sense of computational complexity, but still very
simple in the sense of Kolmogorov complexity: there exist infinite
sequences ${\bf x}$ 
such that for all $n$ there exists short programs
printing $x_{[1:n]}$ ($K(x_{[1:n]}) = O (\log n)$) yet
every such short program takes time exponentially in $n$ \cite[Chapter
7]{LiV97}.
Finally, we 
note that for}
% deterministic infinite sequences ${\bf x}$,
%and 
For certain `special' lengths $n$, we may have $K(x)$
even substantially smaller than $O(\log n)$. For example, suppose $n= 2^m$
for some $m \in \naturals$. Then we can describe $n$ by first
describing $m$ and then describing a program implementing the function
$f(z) = 2^z$. The description of $m$ takes $O(\log m)$ bits, the
description of the program takes a constant number of bits not
depending on $n$. Therefore, for such values of $n$, we get
$K(x) = O( \log m) = O(\log \log n)$.
\paragraph{1(b). Completely Random Objects: $K(x) = n + O(\log n).$}
A {\em code\/} or {\em description method\/} is a binary relation
between source words -- strings to be encoded -- and code words --
encoded versions of these strings. Without loss of generality, we can
take the set of code words to be finite binary strings
\cite{CoverT91}. In this chapter we only consider {\em uniquely
  decodable\/} codes where the relation is one-to-one or one-to-many,
indicating that given an encoding $E(x)$ of string $x$, we can always
reconstruct the original $x$.  The Kolmogorov complexity of $x$ can be
viewed as the code length of $x$ that results from using the {\em
  Kolmogorov code\/} $E^*(x)$: this is the code that encodes $x$ by
the shortest program that prints $x$ and halts.

The following crucial insight will be applied to the Kolmogorov code,
but it is important to realize that in fact it holds for {\em every\/}
uniquely decodable code.  For any uniquely decodable code, there are
no more than $2^m$ strings $x$ which can be described by $m$ bits. The
reason is quite simply that there are no more than $2^m$ binary
strings of length $m$. Thus, the number of strings that can be
described by less than $m$ bits can be at most $2^{m-1} + 2^{m-2}
+ \ldots + 1 < 2^{m}$. In particular, this holds for the code
$E^*$ whose length function is $K(x)$. Thus, the fraction of strings
$x$ of length $n$ with $K(x) < n-k$ is less than $2^{-k}$: the
overwhelming majority of sequences cannot be compressed by more than a
constant.  Specifically, if $x$ is determined by $n$ independent
tosses of a fair coin, then all sequences of length $n$ have the same
probability $2^{-n}$, so that with probability at least $1- 2^{-k}$,
$$K(x) \geq n - k.
$$
On the other hand, for arbitrary $x$, there exists a program `{\tt
  print $x$; halt}'. This program seems to have length $n + O(1)$
where $O(1)$ is a small constant, accounting for the `print' and
`halt' symbols. We have to be careful though: computer programs are
usually represented as a sequence of bytes. Then in the program above
$x$ cannot be an arbitrary sequence of bytes, because we somehow have
to mark the end of $x$. Although we represent both the program and the
string $x$ as bits rather than bytes, the same problem remains. To
avoid it, we have to encode $x$ in a prefix-free manner
(Section~\ref{sec:codingprelim}) which takes $n + O(\log n)$ bits,
rather than $n + O(1)$. Therefore, for all $x$ of length $n$, $ K(x)
\leq n + O(\log n)$. Except for a fraction of $2^{-c}$ of these, $K(x)
\geq n - c$ so that for the overwhelming majority of $x$,
\begin{equation}
\label{eq:inc}
K(x) = n + O(\log n).
\end{equation}
Similarly, if $x$ is determined by independent tosses of a fair coin,
then (\ref{eq:inc}) holds with overwhelming probability. Thus, while
for very regular strings, the Kolmogorov complexity is small
(sublinear in the length of the string), {\em most\/} strings have
Kolmogorov complexity about equal to their own length. Such strings
are called {\em (Kolmogorov)  random\/}: they do not exhibit any
discernible pattern. A more precise definition follows in
Example~\ref{ex:goedel}.
\paragraph{1(c). Stochastic Objects: $K(x) = \alpha n + o(n).$}
Suppose ${\bf x} = x_1 x_2 \ldots$ where the individual $x_i$ are
realizations of some random variable $X_i$, distributed according to
some distribution $P$. For example, we may have that all outcomes
$X_1, X_2, \ldots$ are independently identically distributed (i.i.d.)
with for all $i$, $P(X_i =1) = p$ for some $p \in [0,1]$. In that
case, as will be seen in Section~\ref{sec:universal},
Theorem~\ref{theo.eq.entropy},
\begin{equation}
\label{eq:stochast}
K(x_{[1:n]}) = n \cdot H(p) + o(n),
\end{equation}
where $\log$ is logarithm to the base 2, and $H(p) = - p \log p - (1-p)
\log (1-p)$ is the binary entropy, defined in
Section~\ref{sec:probcode}. For now the important thing to note is
that $0 \leq H(p) \leq 1$, with $H(p)$ achieving its maximum $1$ for
$p = 1/2$. Thus, if data are generated by independent tosses of a fair
coin, (\ref{eq:stochast}) is consistent with (\ref{eq:inc}).  If data
are generated by a biased coin, then the Kolmogorov complexity will
still increase linearly in $n$, but with a factor less than 1 in
front: the data can be compressed by a linear amount. This still holds
if the data are distributed according to some $P$ under which the
different outcomes are dependent, as long as this $P$ is
`nondegenerate'.\footnote{This means that there exists an $\epsilon >
  0$ such that, for all $n \geq 0$, all $x^n \in \{0,1\}^n$, for $a
  \in \{0,1\}$, $P(x_{n+1} = a\mid x_1, \ldots, x_n) > \epsilon$.}  An
example is a {\em $k$-th order Markov chain}, where the probability of
the $i$-th bit being a 1 depends on the value of the previous $k$
bits, but nothing else.  If none of the $2^k$ probabilities needed to
specify such a chain are either $0$ or $1$, then the chain will be
`nondegenerate' in our sense, implying that, with $P$-probability 1,
$K(x_1, \ldots, x_n)$ grows linearly in $n$.

\paragraph{2. Invariance}
It would seem that $K(x)$ depends strongly on what programming
language we used in our definition of $K$. However, it turns out that,
for any two universal languages $L_1$ and $L_2$, letting $K_1$ and
$K_2$ denote the respective complexities, for all $x$ of each length,
\begin{equation}
\label{eq:inva}
|K_1(x) - K_2(x)|  \leq C,
\end{equation}
where $C$ is a constant that depends on $L_1$ and $L_2$ {\em but
  not on $x$ or its length}. Since we allow {\em any\/} universal
language in the definition of $K$, $K(x)$ is only defined up to an
additive constant. This means that the theory is inherently {\em
  asymptotic\/}: it can make meaningful statements pertaining to
strings of increasing length, such as $K(x_{[1:n]}) = f(n) + O(1)$ in
the three examples 1(a), 1(b) and 1(c) above. A statement such as
$K(a) = b$ is not very meaningful.

It is actually very easy to show (\ref{eq:inva}). It is known from the
theory of computation that for
any two universal languages $L_1$ and $L_2$, there exists a compiler,
written in $L_1$, translating programs written in $L_2$ into
equivalent programs written in $L_1$. Thus, let $L_1$ and $L_2$ be two
universal languages, and let $\Lambda$ be a program in $L_1$
implementing a compiler translating from $L_2$ to $L_1$.  For
concreteness, assume $L_1$ is LISP and $L_2$ is Java. Let $(p,y)$ be
the shortest combination of Java program plus input that prints a
given string $x$. Then the LISP program $\Lambda$, when given input
$p$ followed by $y$, will also print $x$ and halt.\footnote{To
  formalize this argument we need to setup the compiler in a way such
  that $p$ and $y$ can be fed to the compiler without any symbols
  in between, but this can be done; see Example~\ref{ex:invariance}.}
It follows that $K_{\mbox{\scriptsize LISP}}(x) \leq l(\Lambda) + l(p)
+ l(y) \leq K_{\mbox{\scriptsize Java}}(x) + O(1)$, where $O(1)$ is
the size of $\Lambda$. By symmetry, we also obtain the opposite
inequality. Repeating the argument for general universal $L_1$ and
$L_2$, (\ref{eq:inva}) follows.

\paragraph{3. Uncomputability}
Unfortunately $K(x)$ is not a recursive function: the Kolmogorov
complexity is not computable in general. This means that there exists
no computer program that, when input an arbitrary string, outputs the
Kolmogorov complexity of that string and then halts. We prove
this fact in Section~\ref{sec:kolmogorov}, Example~\ref{ex:uncomputable}. 
Kolmogorov complexity can be
computably approximated (technically speaking, it is {\em upper
  semicomputable\/} \cite{LiV97}), but not in a
practically useful way: while the approximating algorithm with input
$x$ successively outputs better and better approximations $t_1 \geq
t_2 \geq t_3 \geq \ldots$ to $K(x)$, it is (a) excessively slow, and
(b), it is in general impossible to determine whether the current
approximation $t_i$ is already a good one or not. In the words of
\citeN{BarronC91},  (eventually) ``You know, but you do not know you know''.

Do these properties make the theory irrelevant for practical
applications? Certainly not. The reason is that it is possible to
approximate Kolmogorov complexity after all, in the following, weaker
sense: we take some existing data compression program $C$ (for
example, gzip) that allows every string $x$ to be encoded and decoded
computably and even efficiently. We then approximate $K(x)$ as the
number of bits it takes to encode $x$ using compressor $C$. For many
compressors, one can show that for ``most'' strings $x$ in the set of
all strings of interest, $C(x) \approx K(x)$.  Both {\em universal
  coding\/} \cite{CoverT91} and the {\em Minimum Description Length
  (MDL) Principle} (Section~\ref{sec:mdl}) are, to some extent, based
on such ideas. Universal coding forms the basis of most practical
lossless data compression algorithms, and MDL is a practically
successful method for statistical inference. There is an even closer
connection to the {\em normalized compression distance\/} method, a
practical tool for data similarity analysis that can explicitly be
understood as an approximation of an ``ideal'' but uncomputable method
based on Kolmogorov complexity \cite{CilibrasiV05}.
\section{Overview and Summary} 
\label{sec:overview}
Now that we introduced our main concept, we are ready to give a 
summary of the remainder of the chapter. 
\begin{description}
\item[Section~\ref{sec:kolmogorov}: Kolmogorov Complexity -- Details] We motivate our
  definition of Kolmogorov complexity in
  terms of the theory of computation: the Church--Turing thesis
  implies that our choice of description method, based on universal
  computers, is essentially the only reasonable one. We then introduce some
  basic coding
  theoretic concepts, most notably the so-called {\em prefix-free
    codes\/} that form the basis for our version of Kolmogorov
  complexity. Based on these notions, we give a precise definition of
  Kolmogorov complexity and we  fill in some details
  that were left open in the introduction. 
\item[Section~\ref{sec:shannon}: Shannon vs. Kolmogorov] Here we outline the similarities and
  differences in aim and scope of Shannon's and Kolmogorov's
  information theories. Section~\ref{sec:probcode} reviews
  the {\em entropy}, the central concept in Shannon's theory.
  Although their primary aim is quite different, and they are
  functions defined on different spaces, there is a close relation
  between entropy and Kolmogorov complexity
  (Section~\ref{sec:universal}): if data are distributed according to
  some computable distribution then, roughly, {\em entropy is expected
    Kolmogorov complexity}.
  
  Entropy and Kolmogorov complexity are concerned with information in
  a single object: a random variable (Shannon) or an individual
  sequence (Kolmogorov). Both theories provide a (distinct) notion of
  {\em mutual information\/} that measures the information that {\em
    one object gives about another object}. We introduce and compare
  the two notions in Section~\ref{sec:mutual}.
\end{description}
Entropy, Kolmogorov complexity and mutual information are concerned with
{\em
  lossless\/} description or compression: messages
must be described in such a way that from the description, the
original message can be completely reconstructed. 
Extending the theories to {\em lossy\/}
description or compression enables the formalization of more sophisticated
concepts, such as `meaningful
information' and `useful information'. 
\begin{description}
\item[Section~\ref{sec:meaning}: Meaningful Information, 
Structure Function and
  Learning] The idea of the Kolmogorov Structure Function is to encode
  objects (strings) in two parts: a {\em structural\/} and a {\em
    random\/} part. Intuitively, the `meaning' of the string resides
  in the structural part and the size of the structural part
  quantifies the `meaningful' information in the message. The
  structural part defines a `model' for the string.
  Kolmogorov's structure function approach shows that the meaningful
  information is summarized by the {\em simplest\/} model such that
  the corresponding two-part description is not larger than the
  Kolmogorov complexity of the original string.  Kolmogorov's
  structure function is closely related to J. Rissanen's {\em minimum
    description length principle}, which we
  briefly discuss. This is a practical  theory of
  learning from data that can be viewed as a mathematical
  formalization of Occam's Razor.
\item[Section~\ref{sec:philosophy}: Philosophical Implications]
  Kolmogorov complexity has implications for the foundations of
  several fields, including the foundations of mathematics. The
  consequences are particularly profound for the foundations of {\em
    probability\/} and {\em statistics}. For example, it allows us to
  discern between {\em different forms\/} of randomness, which is
  impossible using standard probability theory.  It provides a precise
  prescription for and justification of the use of Occam's Razor in
  statistics, and leads to the distinction between {\em
    epistemological\/} and {\em metaphysical\/} forms of Occam's
  Razor. We discuss these and other implications for the philosophy of
  information in Section~\ref{sec:philosophy}, which may be read
  without deep knowledge of the technicalities described in
  Sections~\ref{sec:kolmogorov}--\ref{sec:meaning}.
\end{description}
\section{Kolmogorov Complexity: Details}
\label{sec:kolmogorov}
In Section~\ref{sec:essentials} we introduced Kolmogorov complexity
and its main features without paying much attention to either (a)
underlying motivation (why is Kolmogorov complexity a useful measure
of information?) or (b) technical details. In this section, we first
provide a detailed such motivation (Section~\ref{sec:motivation}). We
then (Section~\ref{sec:codingprelim}) provide the technical background knowledge needed for a proper
understanding of the concept. Based on this background knowledge, in
Section~\ref{sec:technical} we
provide a definition of Kolmogorov
complexity directly in terms of Turing machines, equivalent to, but at
the same time more complicated and insightful than the definition we
gave in Section~\ref{sec:koldef}. With the help of this new
definition, we then fill in the gaps left open in
Section~\ref{sec:essentials}.
\subsection{Motivation}
\label{sec:motivation}
Suppose we want to describe a given object by a finite binary string.
We do not care whether the object has many descriptions; however, each
description should describe but one object.  From among all
descriptions of an object we can take the length of the shortest
description as a measure of the object's complexity.  It is natural to
call an object ``simple'' if it has at least one short description,
and to call it ``complex'' if all of its descriptions are long.  But
now we are in danger of falling into the trap so eloquently described
in the Richard-Berry paradox, where we define a natural number as
``the least natural number that cannot be described in less than
twenty words.'' If this number does exist, we have just described it
in thirteen words, contradicting its definitional statement. If such a
number does not exist, then all natural numbers can be described in
fewer than twenty words.  We need to look very carefully at what kind
of descriptions (codes) $D$ we may allow.  If $D$ is known to both a
sender and receiver, then a message $x$ can be transmitted from sender
to receiver by transmitting the description $y$ with $D(y)=x$. We may
define the descriptional complexity of $x$ under specification method
$D$ as the length of the shortest $y$ such that $D(y) = x$.
Obviously, this descriptional complexity of $x$ depends crucially on
$D$: the syntactic framework
of the description language determines the succinctness of
description.  Yet in order to objectively compare descriptional
complexities of objects, to be able to say ``$x$ is more complex than
$z$,'' the descriptional complexity of $x$ should depend on $x$ alone.
This complexity can be viewed as related to a universal description
method that is a priori assumed by all senders and receivers.  This
complexity is optimal if no other description method assigns a lower
complexity to any object.

We are not really interested in optimality with respect to
all description methods.
For specifications to be useful at all it is
necessary that the mapping from $y$ to $D(y)$
can be executed in an effective manner. That is,
it can at least in principle be performed by humans or machines.
This notion has been 
formalized as that of ``partial recursive functions'', 
also known simply as {\em computable\/} functions. 
According to
generally accepted mathematical viewpoints -- the so-called `Church-Turing thesis' --  it coincides
with the intuitive notion of effective computation \cite{LiV97}.
 
The set of partial recursive functions
contains an optimal function that minimizes
description length of every other such function. We denote
this function by $D_0$.
Namely, for any other recursive function $D$,
for all objects $x$,
there is a description $y$ of $x$ under $D_0$ that is
shorter than any description $z$ of $x$ under $D$. (That is,
shorter up to an
additive constant that is independent of $x$.)
Complexity with respect to $D_0$ minorizes 
the complexities with respect
to all partial recursive functions (this is just the invariance result (\ref{eq:inva}) again).

We identify the
length of the description of $x$ with respect
to a fixed specification function $D_0$ with
the ``algorithmic (descriptional) complexity'' of $x$.
The optimality of $D_0$ in the sense above
means that the complexity of an object $x$
is invariant (up to an additive constant
independent of $x$) under transition
from one optimal specification function to another.
Its complexity is an objective attribute
of the described object alone: it is an
intrinsic property of that object, and it does
not depend on the description formalism.
This complexity can be viewed as ``absolute information content'':
the amount of information that needs to be transmitted
between all senders and receivers when they communicate the
message in absence of any other a priori knowledge
that restricts the domain of the message.
This motivates  the program for
a general theory of algorithmic complexity and information.
The 
four
major innovations are as follows:
\begin{enumerate}
\item
In restricting
ourselves to formally effective descriptions,
our definition covers every form of description
that is intuitively acceptable as being effective
according to general viewpoints in mathematics and logic.
\item
The restriction to effective descriptions
entails that there is a universal description
method that minorizes the description length or complexity
with respect to any other effective description
method.
Significantly, this implies Item 3.
\item
The description length or complexity of an object
is an intrinsic attribute of the object independent
of the particular description method or formalizations
thereof.
\item
The disturbing Richard-Berry paradox above does not disappear,
but resurfaces in the form of an alternative
approach to proving G\"odel's  famous
%\index{G\"odel, K.}
result that not every true mathematical statement
is provable in mathematics (Example~\ref{ex:goedel} below).
\end{enumerate}
\subsection{Coding Preliminaries}
\label{sec:codingprelim}
\paragraph{Strings and Natural Numbers}
Let ${\cal X}$ be some finite or countable set. We use the notation
${\cal X}^*$ to denote the set of finite 
{\em strings\/} or {\em sequences\/} over ${\cal X}$. For example,
$$\{0,1\}^* = \{ \epsilon,0,1,00,01,10,11,000,\ldots \},$$
with $\epsilon$ denoting the {\em empty word} `' with no letters.
We identify the natural numbers
${\naturals}$ and $\{0,1\}^*$ according to the
correspondence 
\begin{equation}
\label{eq:correspondence}
(0, \epsilon ), (1,0), (2,1), (3,00), (4,01), \ldots
\end{equation}
The {\em length} $l(x)$ of $x$ is the number of bits
in the binary string $x$. For example,
$l(010)=3$ and $l(\epsilon)=0$. 
If $x$ is interpreted as an integer, we get $ l(x) =  \lfloor \log
(x+1) \rfloor$ and, for $x \geq 2$,
\begin{equation}
\label{eq:intlength}
\lfloor \log x \rfloor
\leq l(x) \leq \lceil \log x \rceil.
\end{equation}
Here, as in the sequel, $\lceil x \rceil$ is the smallest integer larger than or equal to
$x$, $\lfloor x \rfloor$ is the largest integer smaller than or equal
to $x$ and $\log$ denotes logarithm  to base two.
We shall typically be concerned with
encoding finite-length binary strings by other finite-length binary strings.
The emphasis is on binary strings only for convenience;
observations in any alphabet can be so encoded in a way
that is `theory neutral'.

\paragraph*{Codes}
We repeatedly consider the following scenario: a {\em
  sender\/} (say, A) wants to communicate or transmit some information
  to a {\em receiver\/} (say, B). The information to be transmitted is
  an element from some set ${\cal X}$. 
It will be communicated by sending a
binary string, called the {\em message}. 
When B receives the message, he can decode it again and (hopefully)
  reconstruct the element of ${\cal X}$ that was sent.
To achieve this, A and B need to agree
  on a {\em code\/} or {\em description method\/} before
  communicating. Intuitively, this is a binary relation between {\em
  source words} and associated {\em code words}. The relation is fully
  characterized by the {\em decoding function}. Such a decoding function
$D$ can be any function $D: \{ 0, 1 \}^* \rightarrow {\cal X}$.
The domain of $D$ is the set of %
\it code words
\rm and the range of $D$ is the set of %
\it source words. \rm $D(y) = x$ is interpreted as ``$y$ is a code
word for the source word $x$''.
The set of all code words
for source word $x$ is the set $D^{-1} (x) = \{ y: D(y) = x \}$.
Hence, $E=D^{-1}$ can be called the %
\it encoding %
\rm substitution
($E$ is not necessarily a function). With each code $D$ we can
associate a {\em length function\/} $L_D: {\cal X} \rightarrow {\naturals}$ 
such that, for each source
word $x$, $L_D(x)$ is the length of the shortest encoding of $x$:
$$
L_D(x) = \min \{ l(y): D(y) = x  \}.
$$
We denote by $x^*$ the shortest $y$ such that $D(y) = x$; if there is
more than one such $y$, then $x^*$ is defined to be the
first such $y$ in lexicographical order.

In coding theory attention is often restricted to
the case where the source word set is finite, say
${\cal X} =  \{  1, 2,  \ldots , N  \}  $. If there is a constant $l_0$
such that $l(y) = l_0$ for all code words $y$ (equivalently, $L(x) =
l_0$ for all source words $x$),
then we call $D$ a %
\it fixed-length
\rm code. It is
easy to see that $l_0   \geq   \log N$.
For instance, in teletype transmissions the source
has an alphabet of $N = 32$ letters, consisting
of the 26 letters in the Latin alphabet plus
6 special characters. Hence, we need $l_0 = 5$
binary digits per source letter. In electronic computers
we often use the fixed-length ASCII code\index{code!ASCII}
with $l_0=8$.
\paragraph*{Prefix-free code}
In general we cannot uniquely recover $x$
and $y$ from $E(xy)$.  Let $E$ be the identity mapping.  Then we have
$E(00)E(00) = 0000 = E(0)E(000)$.  We now introduce {\em prefix-free
  codes}, which do not suffer from this defect.  A binary string $x$
is a {\em proper prefix} of a binary string $y$ if we can write $y=xz$
for $z \neq \epsilon$.  A set $\{x,y, \ldots \} \subseteq \{0,1\}^*$
is {\em prefix-free} if for any pair of distinct elements in the set
neither is a proper prefix of the other.  A function $D: \{ 0, 1 \}^*
\rightarrow {\naturals}$ defines a {\it prefix-free
  code}\footnote{The standard terminology 
\cite{CoverT91} for such codes is `prefix codes'. Following
\citeN{HarremoesT06}, we use the more informative `prefix-free codes'.}  
if its domain is prefix-free.  In order to decode a code sequence of a
prefix-free code, we simply start at the beginning and decode one code word
at a time. When we come to the end of a code word, we know it is the
end, since no code word is the prefix of any other code word in a
prefix-free code. Clearly, prefix-free codes are uniquely decodable: we can
always unambiguously reconstruct an outcome from its encoding. Prefix
codes are not the only codes with this property; there are uniquely
decodable codes which are not prefix-free. In the next section, we
will define Kolmogorov complexity in terms of prefix-free codes. One
may wonder why we did not opt for general uniquely decodable codes.
There is a good reason for this: It turns out that every uniquely
decodable code can be replaced by a prefix-free code without changing the
set of code-word lengths. This follows from a sophisticated version of
the Kraft inequality \cite[Kraft-McMillan inequality, Theorem
5.5.1]{CoverT91}; the basic Kraft inequality is found in
\cite{HarremoesT06}, Equation 1.1.  In Shannon's and Kolmogorov's
theories, we are only interested in code word {\em lengths\/} of
uniquely decodable codes rather than actual encodings. The
Kraft-McMillan inequality shows that without loss of generality, we
may restrict the set of codes we work with to prefix-free codes, which are
much easier to handle.

\paragraph{Codes for the integers; Pairing Functions}
Suppose we encode each binary string $x=x_1 x_2 \ldots x_n$ as
\[ \bar x = \underbrace{11 \ldots 1}
_{n \mbox{{\scriptsize  \ times}}}0x_1x_2 \ldots x_n .\]
The resulting code is prefix-free because we can determine where the
code word $\bar x$ ends by reading it from left to right without
backing up. Note $l(\bar{x}) = 2n+1$; thus, we have encoded strings in
$\{0,1\}^*$ in a prefix-free  manner at the price of doubling their
length. We can  get a much more efficient code by applying the
construction above to the length $l(x)$ of $x$ rather than $x$ itself:
define 
%\begin{equatiom}
%\label{eq:accent}
$x'=\overline{l(x)}x$, 
%\end{equation}
where $l(x)$ is interpreted as a binary
string according to the correspondence (\ref{eq:correspondence}). Then the 
code that maps $x$ to $x'$ is a prefix-free code satisfying, for all $x \in
\{0,1\}^*$, $l(x') = n+2 \log n+1$ (here we ignore the `rounding
error' in (\ref{eq:intlength})). We call this code the {\em standard
  prefix-free code for the natural numbers} and use $\Lint(x)$ as notation
for the codelength of $x$ under this code: $\Lint(x) = l(x')$. When $x$ is interpreted as a number
(using the correspondence (\ref{eq:correspondence}) and
(\ref{eq:intlength})), we see that $\Lint(x) = \log x + 2 \log \log
x+1$. 

We are often interested in representing a pair of natural numbers (or
binary strings) as a single natural number (binary string). To this
end, we define the {\em standard 1-1 pairing function\/} $\langle \cdot,
\cdot \rangle: \naturals \times \naturals \rightarrow \naturals$ as
$\langle x,y \rangle = x'y$ (in this definition $x$ and $y$ are
interpreted as strings). 
\subsection{Formal Definition of Kolmogorov Complexity}
\label{sec:technical}
In this subsection we provide a formal definition of Kolmogorov
complexity in terms of Turing machines. This will allow us to fill in
some details left open in Section~\ref{sec:essentials}.  Let $T_1 ,T_2
, \ldots$ be a standard enumeration of all Turing machines \cite{LiV97}.
The functions implemented by $T_i$ are called the {\em partial
  recursive} or {\em computable} functions. For technical
reasons, mainly because it simplifies the connection to Shannon's information theory,  we are interested in the so-called prefix complexity, which is
associated with Turing machines for which the set of programs (inputs)
resulting in a halting computation is prefix-free\footnote{There
  exists a version of Kolmogorov complexity corresponding to programs
  that are not necessarily prefix-free, but we will not go into it
  here.}. We can realize this by equipping the Turing machine with a
one-way input tape, a separate work tape, and a one-way output tape.
Such Turing machines are called prefix machines since the halting
programs for any one of them form a prefix-free set.

We first define $K_{T_i}(x)$, the prefix Kolmogorov complexity of $x$
relative to a given prefix machine $T_i$, where $T_i$ is the $i$-th
prefix machine in a standard enumeration of them. $K_{T_i}(x)$ is
defined as the length of the shortest input sequence $y$ such that
$T_i(y) = x$; that is, the $i$-th Turing machine, when run with input $y$, produces $x$ on its output tape and then halts. If no such input sequence exists,
$K_{T_i}(x)$ remains undefined. Of course, this preliminary definition
is still highly sensitive to the particular prefix machine $T_i$ that
we use. But now the `universal prefix machine' comes to our rescue.
Just as there exists universal ordinary Turing machines, there also
exist universal prefix machines. These have the remarkable property
that they can simulate every other prefix machine. More specifically,
there exists a prefix machine $U$ such that, with as input the
concatenation $i'y$ (where $i'$ is the standard encoding of integer
$y$, Section~\ref{sec:codingprelim}), $U$ outputs $T_i(y)$ and then
halts. If $U$ gets  any other input then it does not halt.  
\begin{definition}\label{def.KolmK}
  Let $U$ be our reference prefix machine, i.e. for all $i \in \naturals, y \in \{0,1\}^*$, $U(\langle i, y \rangle) = U(i'y) =
  T_i(y)$. The {\em prefix Kolmogorov complexity} of $x$ is defined
  as $K(x) := K_U(x)$, or equivalently:
\begin{eqnarray}
K(x) & = &
\min_{z} \{ l(z) : U(z) = x , z \in \{0,1\}^*\} =  \nonumber \\
\label{eq:KolmK}
& = &  \min_{i,y}\{l(i') + l(y): T_i (y )=x , y \in \{0,1\}^*, i
\in {\naturals} \}.
\end{eqnarray}
\end{definition}
We can alternatively think of $z$ as a program that prints $x$ and
then halts, or as $z = i'y$ where $y$ is a program such
that, when $T_i$ is input program $y$, it prints $x$ and then halts. 

Thus, by definition $K(x)=l(x^*)$, where $x^*$ is the
lexicographically first shortest self-delimiting (prefix-free) program for
$x$ with respect to the reference prefix machine. Consider the mapping
$E^*$ defined by $E^*(x)=x^*$.  This may be viewed as the encoding
function of a prefix-free code (decoding function) $D^*$ with $D^*(x^*) =
x$. By its definition, $D^*$ is a very parsimonious code.
\begin{example}\label{ex:invariance}
  \rm In Section~\ref{sec:essentials}, we defined $K(x)$ as the
  shortest program for $x$ in some standard programming language such
  as LISP or Java. We now show that this definition is equivalent to
  the prefix Turing machine Definition~\ref{def.KolmK}. Let $L_1$ be a
  universal language; for concreteness, say it is LISP. Denote the
  corresponding Kolmogorov complexity defined as in
  (\ref{eq:kolmogorova}) by $K_{\mbox{\scriptsize LISP}}$.  For the
  universal prefix machine $U$ of Definition~\ref{def.KolmK}, there
  exists a program $p$ in LISP that simulates it \cite{LiV97}. By this
  we mean that, for all $z \in \{0,1\}^*$, either $p(z) = U(z)$ or
  neither $p$ nor $U$ ever halt on input $z$. Run with this program,
  our LISP computer computes the same function as $U$ on its input, so
  that
$$
K_{\mbox{\scriptsize LISP}}(x) \leq l(p) + K_U(x) = K_U(x) +O(1).
$$
On the other hand, LISP, when equipped with the simple input/output
interface described in Section~\ref{sec:essentials}, is a language
such that for all programs $p$, the set of inputs $y$ for which $p(y)$
is well-defined forms a prefix-free set. Also, as is easy to check,
the set of syntactically correct LISP programs is prefix-free.
Therefore, the set of strings $py$ where $p$ is a syntactically
correct LISP program and $y$ is an input on which $p$ halts, is
prefix-free.  Thus we can construct a prefix Turing machine with some
index $i_0$ such that $T_{i_0}(py) = p(y)$ for all $y \in \{0,1\}^*$.
Therefore, the universal machine $U$ satisfies for all $y \in
\{0,1\}^*$, $ U(i_0'py) = T_{i_0}(py) = p(y),$ so that
$$
K_U(x) \leq K_{\mbox{\scriptsize LISP}}(x) + l(i_0') = K_{\mbox{\scriptsize LISP}}(x) + O(1).
$$
We are therefore justified in calling $K_{\mbox{\scriptsize LISP}}(x)$
a version of (prefix) Kolmogorov complexity. 
The same holds for any other universal language, as long as its set of syntactically correct programs is prefix-free. This is the case for every programming language we know of.
\end{example}
\begin{example}{\bf [$K(x)$ as an integer function; uncomputability]}
\label{ex:uncomputable}
\rm 
The correspondence between binary strings and integers established in
(\ref{eq:correspondence}) shows that Kolmogorov complexity may
equivalently be thought of as a function $K: \naturals \rightarrow
\naturals$ where $\naturals$ are the nonnegative integers. This
interpretation is useful to prove that Kolmogorov complexity is
uncomputable.

Indeed, let us assume by means of contradiction that $K$ is
computable. Then the function $\psi(m) := \min_{x \in \naturals} \{ x:
K(x) \geq m \}$ must be computable as well (note that $x$ is
interpreted as an integer in the definition of $\psi$). The definition
of $\psi$ immediately implies $K(\psi(m)) \geq m$. On the other hand,
since $\psi$ is computable, there exists a computer program of some
fixed size $c$ such that, on input $m$, the program outputs $\psi(m)$
and halts. Therefore, since $K(\psi(m))$ is the length of the shortest
program plus input that prints $\psi(m)$, we must have that
$K(\psi(m)) \leq \Lint(m) + c \leq 2 \log m + c$.  Thus, we have $m
\leq 2 \log m + c$ which must be false from some $m$ onwards:
contradiction.
\end{example}
\begin{example}{\bf [G\"odel's incompleteness theorem and randomness]}
\label{ex:goedel}
\rm We say that a formal system (definitions, axioms, rules of inference)
is {\em consistent\/} if no statement which can be expressed in the
system can be proved to be both true and false in the system. A formal
system is {\em sound\/} if only true statements can be proved to be
true in the system. (Hence, a sound formal system is consistent.)

Let $x$ be a finite binary string of length $n$. We write `$x$ is
$c$-random' if $K(x) > n - c$. That is, the shortest binary
description of $x$ has length not much smaller than $x$. We recall
from Section~\ref{sec:key} that the fraction of sequences that can be
compressed by more than $c$ bits is bounded by $2^{-c}$. This shows
that there are sequences which are $c$-random for every $c \geq 1$ and
justifies the terminology: the smaller $c$, the more random $x$.

Now fix any sound formal system $F$ that is powerful enough to express
the statement `$x$ is $c$-random'. Suppose $F$ can be described in $f$
bits. By this we mean that there is a fixed-size program of length $f$
such that, when input the number $i$, outputs a list of all valid
proofs in $F$ of length (number of symbols) $i$. We claim that, for
all but finitely many random strings $x$ and $c \geq 1$, the sentence
`$x$ is $c$-random' is not provable in $F$. Suppose the contrary. Then
given $F$, we can start to exhaustively search for a proof that some
string of length $n \gg f$ is random, and print it when we find such a
string $x$. This procedure to print $x$ of length $n$ uses only $\log
n + f + O(1)$ bits of data, which is much less than $n$. But $x$ is random by
the proof and the fact that $F$ is sound. Hence $F$ is not consistent,
which is a contradiction.
\end{example}
Pushing the idea of Example~\ref{ex:goedel} much further, 
\citeN{Ch87b} proved a particularly strong
variation of G\"odel's theorem, using Kolmogorov complexity but in a more sophisticated way, based on the number $\Omega$ defined below. Roughly, it says the following: there exists an {\em exponential Diophantine equation}, 
\begin{equation}
\label{eq:goedel}
A(n, x_1, \ldots, x_m) = 0
\end{equation}
for some finite $m$, such that the following holds: let $F$ be a
formal theory of arithmetic. Then for all
$F$ that are sound and consistent, there is only a finite number of
values of $n$ for which the theory determines whether
(\ref{eq:goedel}) has finitely or infinitely many solutions $(x_1,
\ldots, x_m)$ ($n$ is to be considered a parameter rather than a
variable).  For all other, infinite number of values for $n$, the
statement `(\ref{eq:goedel}) has a finite number of solutions' is
logically independent of $F$. 
\paragraph{Chaitin's  Number of Wisdom $\Omega$}
An axiom system that can be effectively described by a finite string has 
limited information content -- this was the basis for our proof of G\"odel's theorem above. On the other hand, there exist quite short strings which are
mathematically well-defined but uncomputable, which have an astounding
amount of information in them about the truth of mathematical
statements. Following \citeN{Chaitin75}, we define the {\em halting
  probability\/} $\Omega$ as the real number defined by
$$
\Omega = \sum_{U(p) < \infty} 2^{- l(p)},
$$
the sum taken over all inputs $p$ for which the reference machine
$U$ halts. We call $\Omega$ the
halting probability because it is the probability that $U$ halts if
its program is provided by a sequence of fair coin flips. It turns out
that $\Omega$ represents the {\em halting problem\/} very compactly.
The following theorem is proved in \cite{LiV97}:
\begin{theorem}
  Let $y$ be a binary string of length at most $n$. There exists an
  algorithm A which, given the first $n$ bits of $\Omega$, decides
  whether the universal machine $U$ halts on input $y$; i.e. A outputs
  1 if $U$ halts on $y$; A outputs 0 if $U$ does not halt on $y$; and
  A is guaranteed to run in finite time.
\end{theorem}
The halting problem is a prime example of a problem that is {\em
  undecidable\/} \cite{LiV97}, from which it follows that $\Omega$
must be uncomputable.

Knowing the first 10000 bits of $\Omega$ enables us to solve the
halting of all programs of less than 10000 bits. This includes
programs looking for counterexamples to Goldbach's Conjecture,
Riemann's Hypothesis, and most other conjectures in mathematics which
can be refuted by a single finite counterexample. Moreover, for all
axiomatic mathematical theories which can be expressed compactly
enough to be conceivably interesting to human beings, say in less than
10000 bits, $\Omega_{[1:10000]}$ can be used to decide for every
statement in the theory whether it is true, false, or independent.
Thus, $\Omega$ is truly the number of Wisdom, and `can be known of,
but not known, through human reason' [C.H. Bennett and M. Gardner, {\em
  Scientific American}, 241:11(1979), 20--34]. 
%But even if you possess
%$\Omega_{[1:10000]}$, you cannot use it except by spending time of a
%thoroughly unrealistic nature: the time $t(n)$ it takes to find all
%halting programs of length less than $n$ from $\Omega_{[1:10000]}$
%grows faster than any computable function! \cite{LiV97}
 
%An important side-product of the proof
%is the following insight: although most strings are random in the
%Kolmogorov sense, it is impossible to effectively prove them random.
\subsection{Conditional Kolmogorov complexity}
\label{sec:conditional}
In order to fully develop the theory, we also
need a notion of {\em conditional\/} Kolmogorov complexity.
Intuitively, the conditional  Kolmogorov complexity $K(x|y)$ of
$x$ given $y$ can be interpreted as the shortest program $p$ such
that, when $y$ is given to the program $p$ as input `for free', the
program prints $x$ and then halts. Based on conditional Kolmogorov
complexity, we can then further define Kolmogorov
complexities of more complicated objects such as functions and so on (Example~\ref{ex:kcgeneralobjects}).

The idea of providing $p$ with an input $y$
is realized by putting $\langle y,p \rangle$ rather than just $p$ on the
input tape of a  universal {\em conditional\/} prefix machine $U$.
This is a prefix machine $U$ such that for all $y$, $i$, $q$,
$U(\langle y, \langle i,q \rangle \rangle) = T_i(\langle y, q
\rangle)$, whereas for any input not of this form, $U$ does not halt.
Here $T_1, T_2, \ldots$ is some effective enumeration of prefix
machines. It is easy to show that such a universal conditional prefix
machine $U$ exists \cite{LiV97}.  We now
fix a reference conditional universal prefix machine $U$ and define
$K(x|y)$ as follows:
\begin{definition}\label{def.KolmKb}{\bf [Conditional and Joint Kolmogorov Complexity]}
The {\em conditional prefix Kolmogorov complexity} of $x$ given $y$ (for
free) is
\begin{eqnarray}
K(x|y) & = &  \min_{p}\{l(p): U(\langle y,p \rangle )=x , p \in \{0,1\}^*\}. \\
& = & \min_{q,i}\{l(\langle i, q \rangle) :  U(\langle y, \langle i,q
\rangle \rangle) =x , q \in \{0,1\}^*, i \in
{\naturals} \} \\
& = & \min_{q,i}\{l(i') + l(q) : T_i(y'q) =x , q \in \{0,1\}^*, i \in {\naturals} \}. 
\end{eqnarray}
We define the {\em unconditional complexity\/} $K(x)$ as $K(x)=K(x|\epsilon)$.
We define the {\em joint\/} complexity $K(x,y)$  as $K(x,y) = K(\langle x,y \rangle)$.
\end{definition}
Note that we just redefined $K(x)$ so that the unconditional
Kolmogorov complexity is {\em exactly\/} equal to the conditional
Kolmogorov complexity with empty input. This does not contradict our
earlier definition: having chosen some reference conditional prefix
machine $U$, we can always find an effective enumeration $T'_1, T'_2$
and a corresponding unconditional universal prefix machine $U'$ such
that for all $p$, $U(\langle \epsilon,p \rangle) = U'(p)$. Then we
automatically have, for all $x$, $K_{U'}(x) = K_U(x| \epsilon)$.

\begin{example}{\bf [$K$ for general objects: functions, distributions, sets, ...]}
\label{ex:kcgeneralobjects}
\rm We have defined the Kolmogorov complexity $K$ of binary strings
and natural numbers, which we identified with each other. It is
straightforward to extend the definition to objects such as
real-valued functions, probability distributions and sets. We briefly
indicate how to do this. Intuitively, the Kolmogorov complexity of a
function $f: \naturals \rightarrow \reals$ is the length of the
shortest prefix-free program that computes (outputs) $f(x)$ to
precision $1/q$ on input $x'q'$ for
$q \in \{1,2, \ldots \}$. In terms of conditional universal prefix  machines:
\begin{equation}
\label{eq:kf}
K(f) = \min_{p \in \{0,1\}^*} \bigl\{l(p): \text{for all $q \in \{1,2, \ldots \}, x \in \naturals$:\ }
|U(\langle x, \langle q, p \rangle \rangle) -f(x) | \leq 1/q \bigr\}.
\end{equation}
The Kolmogorov complexity of a function $f: \naturals \times \naturals
\rightarrow \reals$ is defined analogously, with $\langle x, \langle
q, p \rangle \rangle$ replaced by $\langle x, \langle y, \langle q, p
\rangle \rangle \rangle$, and $f(x)$ replaced by $f(x,y)$; similarly
for functions $f: \naturals^k \times \naturals \rightarrow \reals$ for
general $k \in \naturals$.  As a special case of (\ref{eq:kf}), the
Kolmogorov complexity of a probability distribution $P$ is the
shortest program that outputs $P(x)$ to precision $q$ on input
$\langle x,q \rangle$. We will encounter $K(P)$ in
Section~\ref{sec:shannon}.
  
The Kolmogorov complexity of {\em sets\/} can be defined in various
manners \cite{GacsTV01}. In this chapter we only consider finite sets
$S$ consisting of finite strings. One reasonable method of defining
their complexity $K(S)$ is as the length of the shortest program that
sequentially outputs the elements of $S$ (in an arbitrary order) and
then halts. Let $S = \{x_1, \ldots, x_n \}$, and assume that $x_1,
x_2, \ldots, x_n$ reflects the lexicographical order of the elements
of $S$. In terms of conditional
prefix machines, $K(S)$ is the length of the shortest binary program
$p$ such that $U(\langle \epsilon, p \rangle) = z$, where 
\begin{equation}
\label{eq:set}
z = 
\langle x_1, \langle
x_2, \ldots, \langle x_{n-1}, x_n \rangle \ldots \rangle \rangle.
\end{equation} 
This
definition of $K(S)$ will be used in Section~\ref{sec:meaning}. There
we also need the notion of the Kolmogorov complexity of a string $x$
given that $x \in S$, denoted as $K(x | S)$. This is defined as the
length of the shortest binary program $p$ from which the (conditional
universal) $U$ computes $x$ from input $S$ given literally, in the
form of (\ref{eq:set}). 
\end{example}
This concludes our treatment of the basic concepts of Kolmogorov
complexity theory. In the next section we compare these to the basic
concepts of Shannon's information theory.
\section{Shannon and Kolmogorov} 
\label{sec:shannon}
In this section we compare Kolmogorov complexity to Shannon's  [1948] 
information theory, more commonly simply known as `information theory'.
Shannon's theory predates
Kolmogorov's by about 25 years.  Both theories measure the amount of
information in an object as the length of a description of the object.
In the Shannon approach, however, the method of encoding objects is
based on the presupposition that the objects to be encoded are
outcomes of a known random source---it is only the characteristics of
that random source that determine the encoding, not the
characteristics of the objects that are its outcomes.  In the
Kolmogorov complexity approach we consider the individual objects
themselves, in isolation so-to-speak, and the encoding of an object is
a computer program that generates it.  In the Shannon approach we are
interested in the minimum expected number of bits to transmit a
message from a random source of known characteristics through an
error-free channel.  In Kolmogorov complexity we are interested in the
minimum number of bits from which a particular message can effectively
be reconstructed.  A little reflection reveals that this is a great
difference: for {\em every} source emitting but two messages the
Shannon information is at most 1 bit, but we can choose both messages
concerned of arbitrarily high Kolmogorov complexity. Shannon stresses
in his founding article that his notion is only concerned with {\em
  communication}, while Kolmogorov stresses in his founding article
that his notion aims at supplementing the gap left by Shannon theory
concerning the information in individual objects.  To be sure, both
notions are natural: Shannon ignores the object itself but considers
only the characteristics of the random source of which the object is
one of the possible outcomes, while Kolmogorov considers only the
object itself to determine the number of bits in the ultimate
compressed version irrespective of the manner in which the object
arose.

These differences notwithstanding, there exist very strong connections
between both theories. In this section we given an overview of these.
In Section~\ref{sec:probcode} we  recall the relation between
probability distributions and codes, and we review Shannon's fundamental
notion, the {\em entropy}. We then (Section~\ref{sec:lacuna}) indicate
how Kolmogorov complexity resolves a lacuna in the Shannon theory,
namely its inability to deal with information in individual objects.
In Section~\ref{sec:universal} we make precise and explain the
important relation
$$
\text{Entropy} \ \approx \ \text{expected Kolmogorov complexity}.
$$
Section~\ref{sec:mutual} deals with Shannon and algorithmic {\em
  mutual information}, the second fundamental concept in both
theories.
\subsection{Probabilities, Codelengths, Entropy}
\label{sec:probcode}
We now briefly recall the two fundamental relations between
probability distributions and codelength functions, and indicate their
connection to the entropy, the fundamental concept in Shannon's
theory.  These relations are essential for understanding the
connection between Kolmogorov's and Shannon's theory. For (much) more
details, we refer to \citeN{HarremoesT06}'s chapter in this handbook,
and, in a Kolmogorov complexity context, to \cite{GrunwaldV03}. We use
the following notation: let $P$ be a probability distribution defined
on a finite or countable set ${\cal X}$. In the remainder of the
chapter, we denote by $X$ the random variable that takes values in
${\cal X}$; thus $P(X= x) = P(\{x\})$ is the probability that the
event $\{x\}$ obtains. We write $P(x)$ as an abbreviation of $P(X=
x)$, and we write $E_P[f(X)]$ to denote the expectation of a function
$f: {\cal X} \rightarrow \reals$, so that $E_P[f(X)] = \sum_{x \in
  {\cal X}} P(x) f(x)$.

\paragraph*{The Two Relations between probabilities 
and code lengths} 
\begin{enumerate}
\item For every distribution $P$ defined on a finite or countable set
  ${\cal X}$, there exists a code with lengths $L_P(x)$, satisfying,
  for all $x \in {\cal X}$, $L_P(x) = \lceil - \log P(x) \rceil$. This
  is the so-called {\em Shannon-Fano\/} code corresponding to $P$. The
  result follows directly from the Kraft inequality \cite[Section 1.2]{HarremoesT06}.
\item If $X$ is distributed according to $P$, then the Shannon-Fano code corresponding to $P$ is (essentially) the optimal code to use in an expected sense. 
  
  Of course, we may choose to encode outcomes of $X$ using a code
  corresponding to a distribution $Q$, with lengths $\lceil - \log
  Q(x) \rceil$, whereas the outcomes are actually distributed
  according to $P \neq Q$. But, as expressed in the {\em noiseless
    coding theorem\/} or, more abstractly, in \cite[Section 1.3]{HarremoesT06} as
  the {\em First main theorem of information theory}, such a code
  cannot be significantly better, and may in fact be much worse than the code
  with lengths $\lceil - \log P(X) \rceil$: the noiseless coding
  theorem says that
\begin{equation}
\label{eq:nct}
E_P[- \log P(X) ] \leq \min_{C: \ C \text{\ is a prefix-free code}}
E_P[L_C(X)] \leq E_P[- \log P(X) ] + 1, 
\end{equation} 
so that it follows in particular that the expected length of the
Shannon-Fano code satisfies
$$
E_P \lceil- \log P(X) \rceil \leq E_P[- \log P(X) ]  +1 \leq 
\min_{C: \ C \text{\ is a prefix-free code}} E_P[L_C(X)] + 1.
$$ 
and is thus always within just bit of the code that is optimal in
expectation.
\end{enumerate}
In his 1948 paper, Shannon proposed a measure of information in a
distribution, which he called the `entropy', a concept discussed at
length in the chapter by \citeN{HarremoesT06} in this handbook.  It is
equal to the quantity appearing on the left and on the right in
(\ref{eq:nct}):\begin{definition}{\bf [Entropy]} \rm Let ${\cal X}$ be
  a finite or countable set, let $X$ be a random variable taking
  values in ${\cal X}$ with distribution $P$. Then the (Shannon-) \it
  entropy\index{entropy|bold}\index{$H$: entropy stochastic source} %
  \rm of random variable $X$ is given by
\begin{equation}
\label{eq:entropy}
H(P)  =  - \sum_{x \in {\cal X}} P(x) \log P(x) ,
\end{equation}
\end{definition} 
Entropy is defined here as a functional mapping a distribution on
${\cal X}$ to real numbers. In practice, we often deal with a pair of
random variables $(X,Y)$ defined on a joint space ${\cal X} \times
{\cal Y}$. Then $P$ is the joint distribution of $(X,Y)$, and $P_X$ is
its corresponding marginal distribution on $X$, $P_X(x) = \sum_y
P(x,y)$. In that case, rather than writing $H(P_X)$ it is customary to
write $H(X)$; we shall follow this convention below.

Entropy can be interpreted in a number of ways. The noiseless coding
theorem (\ref{eq:nct}) gives a precise coding-theoretic
interpretation: it shows that the entropy of $P$ is essentially equal
to the average code length when encoding an outcome of $P$, if
outcomes are encoded using the optimal code (the code that minimizes
this average code length).
\subsection{A Lacuna in Shannon's Theory}
\label{sec:lacuna}
\begin{example}
\label{ex:00}
\rm
Assuming that $x$ is emitted by a random source $X$
with probability $P(x)$, we can transmit $x$ using the Shannon-Fano
code. This uses (up to rounding) $- \log P(x)$ bits.
By Shannon's noiseless coding theorem this is optimal {\em on average},
the average taken over the probability distribution of outcomes
from the source. Thus, if $x = 00 \ldots 0$ ($n$ zeros), and the
random source emits $n$-bit messages with equal probability $1/2^n$
each, then we require $n$ bits to transmit $x$ (the same as 
transmitting $x$ literally). However, we can transmit $x$ 
in about $\log n$ bits if we ignore probabilities and
just describe $x$ individually. Thus, the optimality with
respect to the average may be very sub-optimal in individual cases. 
\end{example}
In Shannon's theory `information' is fully determined by the
probability distribution on the set of possible messages, and
unrelated to the meaning, structure or content of individual messages.
In many cases this is problematic, since the distribution generating
outcomes may be unknown to the observer or (worse), may not exist at
all\footnote{Even if we adopt a Bayesian (subjective) interpretation
  of probability, this problem remains \cite{Grunwald07}.}. For
example, can we answer a question like ``what is the information in
this book'' by viewing it as an element of a set of possible books
with a probability distribution on it? This seems unlikely.
Kolmogorov complexity provides a measure of information that, unlike
Shannon's, does not rely on (often untenable) probabilistic
assumptions, and that takes into account the phenomenon that `regular'
strings are compressible. Thus, it measures the information content of
an {\em individual finite object.} The fact that such a measure exists
is surprising, and indeed, it comes at a price: unlike Shannon's,
Kolmogorov's measure is asymptotic in nature, and not computable in
general. Still, the resulting theory is closely related to Shannon's,
as we now discuss.
\subsection{Entropy and Expected Kolmogorov Complexity}
\label{sec:universal}
We call a distribution $P$ computable if it can be
  computed by a finite-size program, i.e. if it has
  finite Kolmogorov complexity $K(P)$
  (Example~\ref{ex:kcgeneralobjects}). The set of computable
    distributions is very large: it contains, for example, all Markov
    chains of each order with rational-valued
    parameters. In the following discussion we shall restrict
    ourselves to computable distributions; 
extensions to the uncomputable case are discussed by \citeN{GrunwaldV03}.
%at the end of the section we briefly address the
%    uncomputable case.

If
$X$ is distributed according to some distribution $P$,
then the optimal (in the average sense) code to use is the
Shannon-Fano code. But now suppose it is only known that
$P \in {\cal P}$, where ${\cal P}$ is a large set of computable
distributions, perhaps even the set of all computable
distributions.
Now it is not clear
what code is optimal. We may try the Shannon-Fano code for a particular $P
\in {\cal P}$, but such a code will typically lead to very large
expected code lengths if $X$ turns out to be distributed according to
some $Q \in {\cal P}, Q \neq P$. 
We may ask whether there exists another
code that is `almost' as good as the Shannon-Fano code for $P$, no
matter what $P \in {\cal P}$ actually generates the sequence?
We now show that, (perhaps surprisingly), the answer is yes. 

Let $X$ be a random variable taking on values in the set $\{0,1\}^*$
of binary strings of arbitrary length, and let $P$ be the distribution
of $X$.  $K(x)$ is fixed for each $x$ and gives the shortest code word
length (but only up to a fixed constant). It is {\em independent} of
the probability distribution $P$. Nevertheless, if we weigh each
individual code word length for $x$ with its probability $P(x)$, then
the resulting $P$-expected code word length $\sum_x P(x)K(x)$ almost
achieves the minimal average code word length $ H(P)= - \sum_x P(x)
\log P(x)$.  This is expressed in the following theorem (taken from
\cite{LiV97}):
\begin{theorem}\label{theo.eq.entropy}
  Let $P$ be a computable probability distribution on $\{0,1\}^*$. Then
\[ 0 \leq \left( \sum_x P(x) K(x) - H(P) \right) \leq  K(P) + O(1). \]
\end{theorem}
The theorem becomes interesting if we consider sequences of $P$ that
assign mass to binary strings of increasing length. For example, let
$P_n$ be the distribution on $\{0,1\}^n$ that corresponds to $n$
independent tosses of a coin with bias $q$, where $q$ is computable
(e.g., a rational number). We have $K(P_n) = O(\log n)$, since we can
compute $P_n$ with a program of constant size and input $n,q$ with
length $l(n') + l(q') = O(\log n)$.  On the other hand, $H(P_n) = n
H(P_1)$ increases linearly in $n$ (see, e.g., the chapter by
\citeN{HarremoesT06} in this handbook; see also paragraph 1(c) in
Section~\ref{sec:key} of this chapter). So for large $n$, the optimal
code for $P_n$ requires on average $n H(P_1)$ bits, and the Kolmogorov
code $E^*$ requires only $O(\log n)$ bits extra. Dividing by $n$, we
see that the additional number of bits needed per outcome using the
Kolmogorov code goes to $0$.
%The Shannon-Fano code for a computable distribution is itself
%computable. Therefore, for every computable distribution $P$, the
%universal code $D^*$ whose length function is the Kolmogorov
%complexity compresses on average at least as much as the Shannon-Fano
%code for $P$.  This is the intuitive reason why, no matter what
%computable distribution $P$ we take, its expected Kolmogorov
%complexity is close to its entropy. 
Thus, remarkably, whereas the
entropy is the expected codelength according to $P$ under the optimal
code for $P$ (a code that will be wildly different for different $P$),
there exists a single code (the Kolmogorov code), which is
asymptotically almost optimal for {\em all\/} computable $P$.
\commentout{
\paragraph{Uncountable Sets of Distributions} 
Theorem~\ref{theo.eq.entropy} only applies to computable
distributions.  Nevertheless, in statistics one often deals with
`simple' yet uncountable models such as the Bernoulli model.
According to this model, the data are generated by independent tosses
of a coin with bias $p$, for $p \in [0,1]$. Because $p$ takes values
in an uncountable set and the set of all computable distributions is
countable, Theorem~\ref{theo.eq.entropy} is not directly applicable.
However, as shown by \citeN{BarronC91}, essentially all statistical
models ${\cal M}$ that are used in practice can be approximated by
countable models in such a way that for all $P \in {\cal M}$, for all
$x$ of each length, $K(x) < - \log P(x) + o(n)$, where $o(n)$ is
sublinear (often even logarithmic) in $n$, the length of $x$.  This
fact can be exploited to prove analogues of
Theorem~\ref{theo.eq.entropy} that hold for `statistically
interesting' model classes such as the set of all Bernoulli models,
all Markov chains, and so on. We omit further details.  }
\subsection{Mutual Information}
\label{sec:mutual}
Apart from entropy, the {\em mutual information\/} is perhaps the most
important concept in Shannon's theory. Similarly, apart from
Kolmogorov complexity itself, the {\em algorithmic mutual
  information\/} is one of the most important concepts in Kolmogorov's
theory. In this section we review Shannon's notion, we introduce
Kolmogorov's notion, and then we provide an analogue of
Theorem~\ref{theo.eq.entropy} which says that essentially, Shannon
mutual information is averaged algorithmic mutual information. 
\paragraph{Shannon Mutual Information}
How much information can a random variable $X$ convey about a random
variable $Y$? This is determined by the (Shannon) {\em mutual
  information\/} between $X$ and $Y$. Formally, it is defined as
%We recall its definition below. For more details, including an alternative yet equivalent
%definition, we refer to the chapter by \citeN{HarremoesT06} in this handbook.
%The mutual information between $X$ and $Y$ is defined as
\begin{eqnarray}
\label{eq:mutinf}
I(X; Y) & := &  H(X) - H(X|Y) \\ \nonumber
& = & H(X) + H(Y) - H(X,Y)
\end{eqnarray}
where $H(X|Y)$ is the conditional entropy of $X$ given $Y$, and
$H(X,Y)$ is the joint entropy of $X$ and $Y$; the definition of
$H(X,Y), H(X|Y)$ as well as an alternative but equivalent definition
if $I(X; Y)$, can be found in \cite{HarremoesT06}. The equality between the first and second line follows by straightforward rewriting.
The mutual information can be thought of as the expected (average)
reduction in the number of bits needed to encode $X$, when an outcome
of $Y$ is given for free. In accord with intuition, it is easy to show
that $I(X; Y) \geq 0$, with equality if and only if $X$ and $Y$ are
independent, i.e. $X$ provides no information about $Y$. Moreover, and
less intuitively, a straightforward calculation shows that this
information is {\it symmetric}: $I(X;Y) = I(Y;X)$.

\paragraph{Algorithmic Mutual Information}
In order to define algorithmic mutual information, it will be
convenient to introduce some new notation: We will denote by $\lea$ an
inequality to within an additive constant. More precisely, let $f,g$
be functions from $\{0,1\}^*$ to ${\reals}$. Then by `$f(x) \lea
g(x)$' we mean that there exists a $c$ such that for all $x \in
\{0,1\}^*$, $f(x) < g(x) + c$. We write `$f(x) \gea g(x)$' if $g(x)
\lea f(x)$.  We denote by $\eqa$ the situation when both $\lea$ and
$\gea$ hold.

Since $K(x,y) = K(x'y)$ (Section~\ref{sec:conditional}), trivially, the symmetry property holds: $K(x,y) \eqa K(y,x)$.
An interesting property is the ``Additivity of Complexity''
property
\begin{equation}\label{eq.soi}
  K(x, y) \eqa K(x) + K(y \mid x^*) \eqa K(y) + K(x \mid y^*).
 \end{equation}
 where $x^*$ is the first (in standard enumeration order) shortest
 prefix program that generates $x$ and then halts.  
%It is easy to see
% that $x^*$ has the same information as the pair $x,K(x)$: given $x^*$
% we can compute $x$ and $l(x^*)=K(x)$; given $x,K(x)$ we can run all
% programs simultaneously in dovetailed fashion and select the first
% program of length $K(x)$ that halts with output $x$ as $x^*$.
% (Dovetailed fashion means that in phase $k$ of the process we run all
% programs $i$ for $j$ steps such that $i+j=k$, $k=1,2, \ldots$)
 (\ref{eq.soi}) is the Kolmogorov complexity equivalent of the entropy
 equality $H(X,Y) = H(X) + H(Y|X)$ (see Section I.5 in the chapter by
 \citeN{HarremoesT06}). That this latter equality holds is true by
 simply rewriting both sides of the equation according to the
 definitions of averages of joint and marginal probabilities.  In
 fact, potential individual differences are averaged out.  But in the
 Kolmogorov complexity case we do nothing like that: it is quite
 remarkable that additivity of complexity also holds for individual
 objects.  The result (\ref{eq.soi}) is due to \citeN{Gacs74}, can be
 found as Theorem 3.9.1 in \cite{LiV97} and has a difficult proof.  It
 is perhaps instructive to point out that the version with just $x$
 and $y$ in the conditionals doesn't hold with $\eqa$, but holds up to
 additive logarithmic terms that cannot be eliminated.

To define the algorithmic mutual information between
two individual objects $x$ and $y$ with no
probabilities involved, it is instructive to first recall
the probabilistic notion (\ref{eq:mutinf}). The algorithmic 
definition is, in fact, entirely analogous, with $H$ replaced by $K$ and random variables replaced by individual sequences or their generating programs: 
The
{\em information in  $y$ about $x$}
 is defined as
 \begin{equation}\label{def.mutinf}
   I(y : x) = K(x) - K(x  \mid  y^*) \eqa K(x) + K(y) - K(x, y),
 \end{equation}
where the second equality is a consequence of~(\ref{eq.soi})
and states that this information is symmetric,
$I(x:y) \eqa I(y:x)$, and therefore we can talk about
{\em mutual information}.\footnote{The notation of the
algorithmic (individual)  
 notion $I(x:y)$ distinguishes it from the probabilistic
(average) notion  
$I(X; Y)$.  We deviate slightly from \citeN{LiV97}
where $I(y : x)$ is defined as $K(x) - K(x \mid y)$.}
% \begin{remark}\label{rem.cami} 
%The conditional mutual information is
% \begin{align*}
%   I(x : y \mid z) & = K(x \mid z) - K(x \mid y, K(y \mid z), z)
% \\ & \eqa K(x \mid z) + K(y \mid z) - K(x, y \mid z). 
% \end{align*}
% \end{remark}

Theorem~\ref{theo.eq.entropy} showed that the entropy of distribution
$P$ is approximately equal to the expected (under $P$) Kolmogorov
complexity. Theorem~\ref{thm:mutinf} gives the analogous result for
the mutual information.
\begin{theorem}
\label{thm:mutinf}
Let $P$ be a computable probability distribution on $\{0,1\}^* \times
\{0,1\}^*$. Then 
\begin{equation}\label{eq.eqamipmi}
I(X; Y) - K(P)  \lea  \sum_x \sum_y p(x,y) I(x:y)
\lea I(X;Y) + 2 K(P).
\nonumber
\end{equation}
\end{theorem}
Thus, analogously to Theorem~\ref{theo.eq.entropy}, we see that the
expectation of the algorithmic mutual information $I(x:y)$ is close to
the probabilistic mutual information $I(X; Y)$.

Theorems~\ref{theo.eq.entropy} and~\ref{thm:mutinf} do not stand on
their own: it turns out that just about {\em every\/} concept in
Shannon's theory has an analogue in Kolmogorov's theory, and in all
such cases, these concepts can be related by theorems saying that {\em
  if\/} data are generated probabilistically, then the 
Shannon concept is close to the expectation of
the corresponding Kolmogorov concept. Examples are
the probabilistic vs. the algorithmic sufficient statistics, and the
probabilistic rate-distortion function \cite{CoverT91} vs. the algorithmic Kolmogorov
structure function. The algorithmic sufficient statistic and structure
function are discussed in the next section. For a comparison to their
counterparts in Shannon's theory, we refer to \cite{GrunwaldV05}.
\commentout{
\paragraph*{Outlook: A Lacuna for both theories?}
Entropy, Kolmogorov complexity and mutual
(algorithmic) information are concepts that do not distinguish 
between different {\em   kinds\/} of information (such as `meaningful' and `meaningless'
information). This intuitively important distinction apparently cannot
be formalized in either theory.
%Neither do they say anything about
%forms of communication that go beyond the transmission and 
%subsequent reception and/or observation of a single stream
%of symbols; examples of these would be transmission with feedback, or
%a `dialogue' in the form of questions and answers. 
Such more refined notions can be arrived at by {\em constraining\/}
the description methods with which strings are allowed to be encoded,
and by considering {\em lossy\/} rather than {\em lossless\/}
encoding.  Yet the basic notions entropy, Kolmogorov complexity and
mutual information continue to play a fundamental role. The two most
important developments are {\em rate-distortion theory\/} in the
Shannon setting \cite{CoverT91,GrunwaldV03}, dealing with `useful'
information, and the {\em Kolmogorov structure function\/} in
Kolmogorov's setting, dealing with `meaningful' information
\cite{Kolmogorov74b,Shen83,CoverT91,GacsTV01,VereshchaginV02,Vitanyi02,RissanenT05}.
We explain the latter development below.
}
\section{Meaningful Information}
\label{sec:meaning}
The information contained in an individual
finite object (like a finite binary string) is measured
by its Kolmogorov complexity---the length of the shortest binary program
that computes the object. Such a shortest program contains no redundancy:
every bit is information; but is it meaningful information?
If we flip a fair coin to obtain a finite binary string, then with overwhelming
probability that string constitutes its own shortest program. However,
also with overwhelming probability all the bits in the string are meaningless
information, random noise. On the other hand, let an object
$x$ be a sequence of observations of heavenly bodies. Then $x$
can be described by the binary
string $pd$, where $p$ is the description of
the laws of gravity and the observational
parameter setting, while $d$ accounts for the measurement errors:
we can divide the information in $x$ into
meaningful information $p$ and accidental information $d$.
The main task for statistical inference and learning theory is to
distill the meaningful information present in the data. The question
arises whether it is possible to separate meaningful
information from accidental information, and if so, how.
The essence of the solution to this problem is revealed as follows. As
shown by \citeN{VereshchaginV04}, for all $x \in \{0,1\}^*$, we have
\begin{equation}\label{eq.kcmdl}
K(x)  = \min_{i,p} \{K(i) + l(p) : T_i(p) = x\} + O(1),
\end{equation}
where the minimum is taken over
$p \in \{0,1\}^*$ and $i \in \{1,2, \ldots\}$. 
%I SHOW THIS IN NOTES

To get some intuition why (\ref{eq.kcmdl}) holds, note that the
original definition (\ref{def.KolmK}) expresses that $K(x)$ is the sum
of the description length $\Lint(i)$ of some Turing machine $i$ when
encoded using the standard code for the integers, plus the length of a
program such that $T_i(p) = x$. (\ref{eq.kcmdl}) expresses that the
first term in this sum may be replaced by $K(i)$, i.e. the {\em
  shortest\/} effective description of $i$. It is clear that
(\ref{eq.kcmdl}) is never larger than (\ref{eq:KolmK}) plus some
constant (the size of a computer program implementing the standard
encoding/decoding of integers). The reason why (\ref{eq.kcmdl}) is
also never smaller than (\ref{eq:KolmK}) minus some constant is that
there exists a Turing machine $T_k$ such that, for all $i,p$,
$T_k(i^*p) = T_i(p)$, where $i^*$ is the shortest program that prints
$i$ and then halts, i.e. for all $i, p$, $U(\langle k, i^* p \rangle) =
T_i(p)$ where $U$ is the reference machine used in
Definition~\ref{def.KolmK}. Thus, $K(x)$ is bounded by the constant
length $l(k')$ describing $k$, plus $l(i^*) = K(i)$, plus $l(p)$.

The expression \eqref{eq.kcmdl} shows that we can think of Kolmogorov
complexity as the length of a {\em two-part code}.  This way, $K(x)$
is viewed as the shortest length of a two-part code for $x$, one part
describing a Turing machine $T$, or {\em model}, for the {\em regular}
aspects of $x$, and the second part describing the {\em irregular}
aspects of $x$ in the form of a program $p$ to be interpreted by $T$.
The regular, or ``valuable,'' information in $x$ is constituted by the
bits in the ``model'' while the random or ``useless'' information of
$x$ constitutes the remainder.  This leaves open the crucial question:
How to choose $T$ and $p$ that together describe $x$? In general, many
combinations of $T$ and $p$ are possible, but we want to find a $T$
that describes the meaningful aspects of $x$.  Below we show that this
can be achieved using the {\em Algorithmic Minimum Sufficient
  Statistic}. This theory, built on top of Kolmogorov complexity so to
speak, has its roots in two talks by
Kolmogorov \citeyear{Kolmogorov74a,Kolmogorov74b}. Based on
Kolmogorov's remarks, the theory has been further developed by several
authors, culminating in \citeN{VereshchaginV04}, some
of the key ideas of which
we outline below.
\paragraph{Data and Model}
\index{model}
We restrict attention to the following setting: we observe data $x$ in
the form of a finite binary string of some length $n$.  As models for
the data, we consider finite sets $S$ that contain $x$.  
In statistics and machine learning, the use of finite sets is
nonstandard: one usually models the data using probability
distributions or functions.  However, the restriction of sets is just
a matter of convenience: the theory we are about to present
generalizes straightforwardly to the case where the models are
arbitrary computable probability density functions and, in fact, other
model classes such as computable functions \cite{VereshchaginV04}; see also
Section~\ref{sec:mdl}.

The intuition behind the idea of a set as a model is the following:
informally, `$S$ is a good model for $x$' or equivalently, $S$
captures all structure in $x$, if, in a sense to be made precise
further below, it summarizes all simple
properties of $x$. 
In Section~\ref{sec:algsuf} below, we work towards the definition of the
algorithmic minimal sufficient statistic (AMSS) via the
fundamental notions of `typicality' of data and `optimality' of a set.
Section~\ref{sec:ksf} investigates the AMSS further in terms of the
important {\em Kolmogorov Structure Function}. In
Section~\ref{sec:mdl}, we relate the AMSS to the more well-known {\em
  Minimum Description Length Principle}.
\subsection{Algorithmic Sufficient Statistic}
\label{sec:algsuf}
We are now about to formulate the central notions `$x$ is typical for
$S$' and `$S$ is optimal for $x$'. Both are necessary, but not
sufficient requirements for $S$ to precisely capture the
`meaningful information' in $x$. After having introduced optimal sets,
we investigate what further requirements we need.  The development
will make heavy use of the Kolmogorov complexity of sets, and
conditioned on sets. These notions, written as $K(S)$ and $K(x|S)$,
where defined in Example~\ref{ex:kcgeneralobjects}.  
\subsubsection{Typical Elements}
\index{typical data}\index{random data}
\index{data!typical} \index{data!random}
Consider a string $x$
of length $n$ and prefix complexity $K(x)=k$.
We look for the {\em structure} or {\em regularity} in $x$ that is
to be summarized with a set $S$
of which $x$ is a {\em random} or  {\em typical} member:
given $S$ containing $x$, %(or rather,
%shortest program $S^*$ for $S$), 
the element $x$ cannot
be described significantly shorter than by its maximal length index in $S$,
that is, $ K(x \mid S) \geq \log |S| +O(1) $. Formally,

\begin{definition}
\rm
Let $\beta \ge 0$ be an agreed-upon, fixed, constant.
A finite binary string $x$
is a {\em typical} or {\em random} element of a set $S$ of finite binary
strings, if $x \in S$ and
\begin{equation}\label{eq.deftyp}
 K(x \mid S) \ge \log |S| - \beta.
\end{equation}
%where $S^*$ is a shortest program for $S$.
We will not indicate the dependence on $\beta$ explicitly, but the
constants in all our inequalities ($O(1)$) will be allowed to be functions
of this $\beta$.
\end{definition}

This definition requires a finite $S$.
\commentout{
In fact, since
$K(x \mid S) \leq K(x)+O(1) $, it limits the size of $S$ to $O(2^k)$.
%The shortest program $S^*$ from
%which $S$ can be computed is an {\em algorithmic statistic} for $x$ if
%\index{algorithmic statistic}
%\begin{equation}\label{eq.typ}
 %K(x \mid S) \geq \log |S| +O(1).
%\end{equation}
}
Note that the notion of typicality is not absolute
but depends on fixing the constant implicit in the $O$-notation.

\begin{example}\label{xmp.typical}
\rm
% erin laten staan want is een goede uitdrukking voor hoe een set $S$
% als model kan dienen!
Consider the set $S$ of binary strings of length $n$
whose every odd position is 0.
Let $x$ be an element of this set in which the subsequence of bits in
even positions is an incompressible string.
Then $x$ is a typical element of $S$.
But $x$ is also a typical element of the set $\{x\}$.
\end{example}
Note that, if $x$ is not a typical element of $S$, then $S$ is
certainly not a `good model' for $x$ in the intuitive sense described above: $S$ does not capture all
regularity in $x$. However, the example above ($S = \{x \}$) shows that even
if $x$ is typical for $S$, $S$ may still not capture `all meaningful
information in $x$'.

\begin{example} \rm
  If $y$ is not a typical element of $S$, this means that it has some
  simple special property that singles it out from the vast majority
  of elements in $S$. This can actually be proven formally
  \cite{Vitanyi05}. Here we merely give an example.  Let $S$ be as in
  Example~\ref{xmp.typical}. Let $y$ be an element of $S$ in which the
  subsequence of bits in even positions contains two times as many 1s
  than 0s. Then $y$ is not a typical element of $S$: the overwhelming
  majority of elements of $S$ have about equally many 0s as 1s in even
  positions (this follows by simple combinatorics). As shown in
  \cite{Vitanyi05}, this implies that $K(y|S) \ll | \log S|$, so that
  $y$ is not typical.
\end{example} 
\subsubsection{Optimal Sets}
\label{sec:optimal}
\index{optimal model}
\index{model!optimal}
Let $x$ be a binary data string of length $n$.
For every finite set $S \ni x$, we have
$K(x) \leq K(S) + \log |S| + O(1)$,
since we can describe $x$ by giving $S$ and the index of $x$
in a standard enumeration of $S$. Clearly this can be implemented
by a Turing machine computing the finite set $S$ and a program
$p$ giving the index of $x$ in $S$.
The size of a set containing $x$ measures intuitively the number of
properties of $x$ that are represented: 
The largest set is $\{0,1\}^{n}$ and represents only one property
of $x$, namely, being of length $n$. It clearly ``underfits''
as explanation or model for $x$. The smallest set containing $x$
is the singleton set $\{x\}$ and represents all conceivable properties
of $x$. It clearly ``overfits'' as explanation or model for $x$.  

There are two natural measures of suitability of such a set as
a model for $x$.
We might prefer either (a) the simplest set, or (b) the smallest set, as
corresponding to the most likely structure `explaining' $x$.
Both the largest set $\{0,1\}^n$ [having low complexity of about $K(n)$]
and the  singleton set $\{x\}$ [having high complexity of about $K(x)$], 
while certainly statistics for $x$,
would indeed be considered poor explanations.
We would like to balance simplicity of model vs. size of model.
Both measures relate to the optimality of a two-stage description of
$x$ using a finite set $S$ that contains it. Elaborating on
the two-part code described above,
\begin{align}\label{eq.twostage}
 K(x) & \leq  K(S) + K(x \mid S) +O(1)
\\ & \leq K(S) + \log |S| +O(1),
\nonumber
\end{align}
where the first inequality follows because there exists a program $p$
producing $x$ that first computes $S$ and then computes $x$ based on
$S$; if $p$ is not the shortest program generating $x$, then the
inequality is strict.  The second substitution of $K(x \mid S)$ by
$\log |S|+O(1)$ uses the fact that $x$ is an element of $S$.  The
closer the right-hand side of \eqref{eq.twostage} gets to the
left-hand side, the better the two-stage description of $x$ is.  This implies a
trade-off between meaningful model information, $K(S)$, and meaningless
``noise'' $\log |S|$.  A set $S$ (containing $x$) for which
\eqref{eq.twostage} holds with equality,
\begin{equation}\label{eq.optim}
K(x) = K(S) + \log |S| +O(1),
\end{equation}
is called {\em optimal}.  The first line of (\ref{eq.twostage})
implies that if a set $S$ is optimal for $x$, then $x$ must be a
typical element of $S$. However, the converse does not hold: a data
string $x$ can be typical for a set $S$ without that set $S$
being optimal for $x$. 
\begin{example}
\rm 
It can be shown that the set $S$ of Example~\ref{xmp.typical} is also
optimal, and so is $\{x\}$.
Sets for which $x$ is typical form a much wider class than optimal 
sets for $x$: the set
$\{x,y\}$ is still typical for
$x$ but with most $y$ it will be too complex to be optimal for $x$.
A less artificial example can be found in \cite{VereshchaginV04}.
\end{example}
While `optimality' is a refinement of
`typicality', the fact that $\{x \}$ is still an optimal set for $x$
shows that it is still not sufficient by itself to capture the notion 
of `meaningful information'. In order to discuss the necessary
refinement, we first need to connect optimal sets to the notion of a
`sufficient statistic', which, as its name suggests, has its roots in
the statistical literature. 
%\begin{example}
%\rm
%Combining \eqref{eq.twostage} and
%\eqref{eq.optim}, we see that if $S$ is an optimal set for $x$
%then $K(x,S)=K(x)+O(1)$ which implies that $K(S\mid x)=O(1)$.
%Going from $x$ to $S$ requires but an $O(1)$ length program,
%which implies that there are ony $O(1)$ optimal sets for $x$,
%however large $x$ may be.
%\end{example}
\subsubsection{Algorithmic Sufficient Statistic}
A {\em statistic} of the data $x=x_1 \ldots x_n$ is a function $f(x)$.
Essentially, every function will do.  For example, $f_1(x)=n$, $f_2
(x)= \sum_{i=1}^n x_i$, $f_3(x)=n-f_2(x)$, and $f_4 (x)= f_2(x)/n$,
are statistics.  A ``sufficient'' statistic of the data contains all
information in the data about the model.  In introducing the notion of
sufficiency in classical statistics, Fisher [1922] stated: ``The
statistic chosen should summarize the whole of the relevant
information supplied by the sample. This may be called the Criterion
of Sufficiency $\ldots$ In the case of the normal 
distributions it is evident that the second moment is a sufficient
statistic for estimating the standard deviation.''  For example, in
the Bernoulli model (repeated coin flips with outcomes 0 and 1
according to fixed bias), the statistic $f_4$ is sufficient. It gives
the mean of the outcomes and estimates the bias of the Bernoulli
process, which is the only relevant model information.  For the
classic (probabilistic) theory see, for example, \cite{CoverT91}. 
\citeN{GacsTV01} develop an algorithmic theory of sufficient statistics
(relating individual data to individual model) and establish its
relation to the probabilistic version; this work is extended by \citeN{GrunwaldV05}.  The algorithmic
basics are as follows: Intuitively, a model expresses the essence of
the data if the two-part code describing the data consisting of the
model and the data-to-model code is as concise as the best one-part
description.  In other words, we call a shortest program for an
optimal set with respect to $x$ an {\em algorithmic sufficient
  statistic} for $x$.  
%Among optimal sets, there is a direct tradeoff
%between complexity and log size, which together sum to $ K(x)+O(1)$.
\begin{example}
\rm
{\bf (Sufficient Statistic)}
\label{ex:suffragette}
Let us look at a coin toss example.  Let $k$ be a number in the range
$0,1,\dots,n$ of complexity $\log n+ O(1)$ given $n$ and let $x$ be a
string of length $n$ having $k$ 1s of complexity $K(x \mid n,k) \geq
\log {n \choose k}$ given $n,k$. This $x$ can be viewed as a typical
result of tossing a coin with a bias about $p=k/n$.  A two-part
description of $x$ is given by first specifying the number $k$ of 1s
in $x$, followed by the index $j \leq \log |S|$ of $x$ in the set $S$
of strings of length $n$ with $k$ 1s.  This set is optimal, since, to
within $O(1)$, $K(x) = K(x, \langle n,k \rangle) = K(n,k) + K(x \mid
n,k) = K(S)+ \log|S|$. The shortest program for $S$, which amounts to
an encoding of $n$ and then $k$ given $n$, is an algorithmic
sufficient statistic for $x$.
\end{example}
%\begin{example}\label{ex.structure}
%\rm
%{\bf (Hierarchy of Sufficient Statistics)}
%A picture such as the `Mona Lisa', to borrow an example from
%\cite{CoverT91} can be represented as an array of pixels, which can in turn be represented as a binary string. We may be interested in 
%the image depicted, but also in the underlying color pattern
%or pattern of brush strokes. Each of these aspects suggest
%that there is a particular ``model level'' at which there
%is a sufficient statistic for that aspect. An expert trying
%to attribute a painting may aim at finding sufficient statistics
%for many such aspects. This more complex example is formalized by \citeN{Vitanyi05}.
%\end{example}
The optimal set that admits the shortest possible program (or rather
that shortest program) is called {\em algorithmic minimal sufficient
  statistic\/} of $x$. In general there can be more than one such set
and corresponding program:
\begin{definition}[Algorithmic  minimal sufficient statistic]
\label{def:algsufstat}
An {\em algorithmic sufficient statistic\/} of $x$ is a shortest
program for a set $S$ containing $x$ that is optimal, i.e. it
satisfies (\ref{eq.optim}).  An algorithmic sufficient statistic with
optimal set $S$ is {\em minimal\/} if there exists no optimal set $S'$
with $K(S') < K(S)$.
\end{definition}
The algorithmic minimal sufficient statistic (AMSS) divides the
information in $x$ in a relevant structure expressed by the set $S$,
and the remaining randomness with respect to that structure, expressed
by $x$'s index in $S$ of $\log |S|$ bits. The shortest program for $S$
is itself alone an algorithmic definition of structure, without a
probabilistic interpretation.
\begin{example}
\rm {\bf (Example~\ref{xmp.typical}, Cont.)}
The shortest program for the set $S$ of Example~\ref{xmp.typical} is a
minimum sufficient statistic for the string $x$ mentioned in that
example. The program generating the set $\{ x \}$, while still an
algorithmic sufficient statistic, is not a minimal sufficient statistic.
\end{example}
\begin{example}
\rm {\bf (Example~\ref{ex:suffragette}, Cont.)}
The $S$ of Example~\ref{ex:suffragette} encodes the number of $1$s in $x$. The shortest program
for $S$  is an
algorithmic minimal sufficient statistic for {\em most\/} $x$ of
length $n$ with $k$ $1$'s, since only a fraction of at most $2^{-m}$
$x$'s of length $n$ with $k$ $1$s can have $K(x) < \log | S| - m$
(Section~\ref{sec:kolmogorov}). But of course there exist $x$'s with
$k$ ones which have much more regularity. An example is the string
starting with $k$ $1$'s followed by $n-k$ $0$'s. For such strings, $S$ is
not optimal anymore, nor is $S$ an algorithmic sufficient statistic.
\end{example}
To analyze the minimal sufficient statistic further, it is useful to
place a constraint on the maximum complexity of set $K(S)$, say $K(S)
\leq \alpha$, and to
investigate what happens if we vary $\alpha$. The result is the {\em
  Kolmogorov Structure Function}, which we now discuss.
\subsection{The Kolmogorov Structure Function}
\label{sec:ksf}
\begin{figure}
\begin{center}
\epsfxsize=8cm
\epsfxsize=8cm \epsfbox{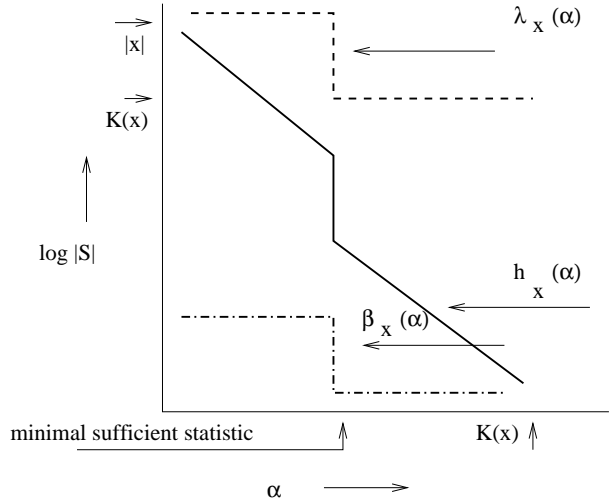}
\end{center}
\caption{Structure functions $h_x(i), \beta_x(\alpha), \lambda_x(\alpha)$,
and minimal sufficient statistic.}
\label{figure.estimator}
\end{figure}

The {\em
  Kolmogorov structure} function \cite{Kolmogorov74a,Kolmogorov74b,VereshchaginV04} $h_x$ of given data $x$ is defined by
 \begin{equation}\label{eq2}
   h_{x}(\alpha) = \min_{S} \{\log | S| : S \ni x,\; K(S) \leq \alpha\},
\end{equation}
where $S \ni x$ is
a contemplated model for $x$, and $\alpha$ is a non-negative
integer value bounding the complexity of the contemplated $S$'s.
Clearly, the Kolmogorov structure function is
nonincreasing and reaches $\log |\{x\}| = 0$
for $\alpha = K(x)+c_1$ where $c_1$ is the number of bits required
to change $x$ into $\{x\}$.
For every $S\ni x$ we have \eqref{eq.twostage}, and
hence $K(x)\le\alpha+h_x(\alpha)+O(1)$; that is, the
function $h_x(\alpha)$
never decreases
more than a fixed independent constant below
the diagonal \emph{sufficiency line} $L$ defined by
$L(\alpha)+\alpha = K(x)$,
which is a lower bound on $h_x (\alpha)$
and is approached to within a constant distance by
the graph of $h_x$ for certain $\alpha$'s
(e.g., for $\alpha = K(x)+c_1$).
For these $\alpha$'s we
thus have
$\alpha + h_x (\alpha) = K(x)+O(1)$;
a model corresponding to such an $\alpha$ (witness for
$h_x(\alpha)$) is a sufficient statistic, and it is
{\em minimal} for the least such $\alpha$ \cite{CoverT91,GacsTV01}.
This is depicted in Figure~\ref{figure.estimator}. Note once again
that the structure function is
defined relative to given data (a single sequence $x$); different
sequences result in different structure functions. Yet, all these
different
functions share some properties: for all $x$, the function
$h_x(\alpha)$ will lie above the diagonal sufficiency line for all
$\alpha \leq \alpha_x$. Here $\alpha_x$ is the complexity $K(S)$ of
the AMSS for $x$. For $\alpha \geq \alpha_x$, the function
$h_x(\alpha)$ remains within a constant of the diagonal. For
stochastic strings generated by a simple computable distribution
(finite $K(P)$), the sufficiency line will typically be first hit for $\alpha$ close to
$0$, since the AMSS will grow as $O(\log n)$. For example, if $x$ is generated by independent fair coin
flips, then, with probability 1, one AMSS will be $S = \{0,1\}^n$ with
complexity $K(S) = K(n) = O(\log n)$.  
One may suspect
that all intuitively `random' sequences have a small sufficient
statistic of order $O(\log n)$ or smaller. Surprisingly, this turns
out not to be the case, as we show in Example~\ref{ex:posneg}.

\begin{example}
  \rm {\bf (Lossy Compression)} \index{compression!lossy} The
  Kolmogorov structure function $h_x( \alpha)$ is relevant to lossy
  compression (used, e.g., to compress images).  Assume we need to
  compress $x$ to $\alpha$ bits where $\alpha\ll K(x)$.  Of course
  this implies some loss of information present in $x$.  One way to
  select redundant information to discard is as follows: let $S_0$ be
  the set generated by the Algorithmic Minimum Sufficient Statistic
  $S_0^*$ ($S_0^*$ is a shortest program that prints $S_0$ and halts).
  Assume that $l(S_0^*) = K(S_0) \leq \alpha$. Since $S_0$ is an
  optimal set, it is also a typical set, so that $K(x|S_0) \approx
  \log |S_0|$. We compress $x$ by $S_0^*$, taking $\alpha$ bits. To
  reconstruct an $x'$ close to $x$, a decompressor can first
  reconstruct the set $S_0$, and then select an element $x'$ of $S_0$
  uniformly at random.  This ensures that with very high probability
  $x'$ is itself also a typical element of $S_0$, so it has the same
  properties that $x$ has. Therefore, $x'$ should serve the purpose of
  the message $x$ as well as does $x$ itself.  However, if $l(S_0^*) >
  \alpha$, then it is not possible to compress all meaningful
  information of $x$ into $\alpha$ bits. We may instead encode, among
  all sets $S$ with $K(S) \leq \alpha$, the one with the smallest
  $\log |S|$, achieving $h_x(\alpha)$. But inevitably, this set will
  not capture all the structural properties of $x$.

Let us look at an example. To transmit a picture of
``rain'' through a channel with limited capacity $\alpha$,
one can transmit the indication that this is a picture of the rain and
the particular drops may be chosen by the receiver at random.
In this interpretation, the complexity constraint $\alpha$ determines
how ``random'' or ``typical'' $x$ will be with respect to the chosen
set $S$ ---and hence how ``indistinguishable'' from the
original $x$ the
randomly reconstructed $x'$ can be expected to be.
\end{example}
We end this section with an example of a strange consequence of 
Kolmogorov's theory:
\begin{example}
\rm
{\bf ``Positive'' and ``Negative'' Individual Randomness:}
\label{ex:posneg}
\citeN{GacsTV01} showed the existence
of strings for which essentially
the singleton set consisting of the string itself is a minimal
sufficient statistic (Section~\ref{sec:algsuf}). While a sufficient
statistic of an object yields a two-part code that is as short as the shortest
one part code, restricting the complexity of the allowed statistic
may yield two-part codes that are considerably longer than the best one-part
code (so that the statistic is insufficient).
In fact,
for every object there is a complexity bound below which this happens;
this is just the point where the Kolmogorov structure function hits
the diagonal. If that bound is small (logarithmic) we call the object ``stochastic''
since it has a simple satisfactory explanation (sufficient statistic).
Thus, Kolmogorov \citeyear{Kolmogorov74a} 
 makes the important distinction of
an object being random in the ``negative'' sense by having this bound
high (it has high complexity and is not a typical element of
a low-complexity model),
and an object being random in the ``positive,
probabilistic'' sense by both having this bound small and itself
having complexity considerably exceeding this bound
(like a string $x$ of length $n$ with $K(x) \geq n$,
being typical for the
set $\{0,1\}^n$, or the uniform probability distribution over that
set,
while this set or probability distribution
has complexity $K(n)+O(1) = O(\log n)$).
We depict the distinction in Figure~\ref{figure.pos_negrandom}.
\begin{figure}
\begin{center}
\epsfxsize=8cm
\epsfxsize=8cm \epsfbox{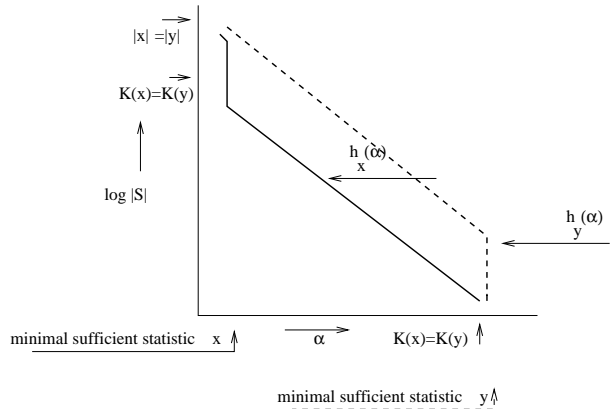}
\end{center}
\caption{Data string $x$ is ``positive random'' or ``stochastic''
 and data string $y$
is just ``negative random'' or ``non-stochastic''.}
\label{figure.pos_negrandom}
\end{figure}
\end{example}
\subsection{The Minimum Description Length Principle}
\label{sec:mdl}
\paragraph{Learning}
The main goal of statistics and machine learning is to learn from
data. One common way of interpreting `learning' is as a search for the
structural, regular properties of the data -- all the patterns that
occur in it. On a very abstract level, this is just what is achieved
by the AMSS, which can thus be related to learning, or, more
generally, inductive inference. There is however another, much more
well-known method for learning based on data compression. This is the
Minimum Description Length (MDL) Principle, mostly developed by J.
Rissanen \citeyear{Rissanen78,Rissanen89} -- see \cite{Grunwald07} for
a recent introduction; see also \cite{Wallace05} for the related MML
Principle. Rissanen took Kolmogorov complexity as an informal starting
point, but was not aware of the AMSS when he developed the first, and,
with hindsight, somewhat crude version of MDL \cite{Rissanen78}, which
roughly says that the best theory to explain given data $x$ is the one
that minimizes the sum of
\begin{enumerate}
\item The length, in bits, of  the description of the theory, plus
\item The length, in bits, of the description of the data $x$ when the
  data is described with the help of the theory.
\end{enumerate}
Thus, data is encoded by first encoding a theory (constituting the
`structural' part of the data) and then encoding the data using the
properties of the data that are prescribed by the theory.  Picking the
theory minimizing the total description length leads to an automatic
trade-off between complexity of the chosen theory and its goodness of
fit on the data. This provides a principle of inductive inference that
may be viewed as a mathematical formalization of `Occam's Razor'.
It automatically protects against overfitting, a central concern of
statistics: when allowing models of arbitrary complexity, we are
always in danger that we model random fluctuations rather than the trend
in the data \cite{Grunwald07}.

The MDL Principle has been designed so as to be practically
useful. This means that the codes used to describe a `theory' are not
based on Kolmogorov complexity. However, there exists an `ideal'
version of MDL \cite{LiV97,BarronC91}  which does rely on Kolmogorov
complexity. Within our framework (binary data, models as sets), it
becomes \cite{VereshchaginV04,Vitanyi05}: pick a set $S \ni x$ minimizing
the two-part codelength
\begin{equation}
K(S) - \log | S|.
\end{equation}
In other words: any ``optimal set'' (as defined in Section~\ref{sec:optimal}) is regarded as a good explanation of
the theory. It follows that every set $S$ that
is an AMSS also minimizes the two-part codelength to within $O(1)$. 
However, as we already indicated, there exist optimal sets $S$ (that,
because of their optimality, may be selected by MDL), that are not
minimal sufficient statistics. As explained by \citeN{Vitanyi05},
these do not capture the idea of `summarizing all structure in
$x$'. Thus, the AMSS may be considered a refinement of the idealized
MDL approach. 
\paragraph{Practical MDL}
The practical MDL approach uses probability distributions rather than
sets as models. Typically, one restricts to distributions in some
model class such as the set of all Markov chain distributions of each
order, or the set of all polynomials $f$ of each degree, where $f$
expresses that $Y = f(X) + Z$, and $Z$ is some normally distributed
noise variable (this makes $f$ a `probabilistic' hypothesis). These
model classes are still `large' in that they cannot be described by a finite
number of parameters; but they are simple enough so that admit
efficiently computable versions of MDL -- unlike the ideal version
above which, because it involves Kolmogorov complexity, is
uncomputable. The Kolmogorov complexity, set-based theory has to be
adjusted at various places to deal with such practical models, one
reason being that they have uncountably many elements. MDL has been
successful in practical statistical and machine learning problems
where overfitting is a real concern \cite{Grunwald07}. Technically, MDL
algorithms are very similar to the popular Bayesian methods, but the
underlying philosophy is very different: MDL is based on finding
structure in {\em individual data\/} sequences; distributions (models)
are viewed as {\em representation languages for expressing useful
  properties of the data\/}; they are neither viewed as objectively
existing but unobservable objects according to which data are
`generated'; nor are they viewed as representing subjective degrees of
belief, as in a mainstream Bayesian interpretation.

In recent years, ever more sophisticated refinements of the original
MDL have developed \cite{Rissanen96,RissanenT05,Grunwald07}. For
example, in modern MDL approaches, one uses {\em universal codes\/}
which may be two-part, but in practice are often {\em one-part\/}
codes.

\section{Philosophical Implications and Conclusion}
\label{sec:philosophy}
We have given an overview of algorithmic information theory, focusing
on some of its most important aspects: Kolmogorov complexity,
algorithmic mutual information, their relations to entropy and Shannon
mutual information, the Algorithmic Minimal Sufficient Statistic and
the Kolmogorov Structure Function, and their relation to `meaningful
information'. Throughout the chapter we emphasized insights that, in
our view, are `philosophical' in nature. It is now time to harvest and
make the philosophical connections explicit. Below we first discuss
some implications of algorithmic information theory on the philosophy
of (general) mathematics, probability theory and statistics. We then
end the chapter by discussing the philosophical implications for
`information' itself.  As we shall see, it turns out
that nearly all of these philosophical implications are somehow
related to {\em randomness}.
\paragraph{Philosophy of Mathematics: Randomness in Mathematics}
In and after Example~\ref{ex:goedel} we indicated that the ideas
behind Kolmogorov complexity are intimately related to G\"odel's  
incompleteness theorem. The finite Kolmogorov complexity of any
effective axiom system implied the existence of bizarre equations like
(\ref{eq:goedel}), whose full solution is, in a sense, random: {\em
  no\/} effective axiom system can fully determine the solutions of
this single equation. In this context,
Chaitin writes: ``This is a region in which mathematical truth has
no discernible structure or pattern and appears to be completely
random [...] Quantum physics has shown that there is randomness in
nature. I believe that we have demonstrated [...] that randomness is
already present in pure Mathematics. This does not mean that the
universe and Mathematics are completely lawless, it means that laws of
a different kind apply: statistical laws. [...] Perhaps number theory
should be pursued more openly in the spirit of an experimental
science!''.
\paragraph{Philosophy of Probability: Individual Randomness}
The statement `$x$ is a random sequence' is essentially meaningless in
classical probability theory, which can only make statements that hold
for ensembles, such as `relative frequencies converge to probabilities
{\em with high probability}, or {\em with probability 1}'. But in
reality we only observe one sequence. What then does the statement
`this sequence is a typical outcome of distribution $P$' or,
equivalently, is `random with respect to $P$' tell us about the
sequence?  We might think that it means that the sequence satisfies
all properties that hold with $P$-probability 1. But this will not
work: if we identify a `property' with the set of sequences satisfying
it, then it is easy to see that the intersection of all sets
corresponding to properties that hold `with probability 1' is empty.
The Martin-L\"of theory of randomness \cite{LiV97} essentially
resolves this issue.  Martin-L\"of's notion of randomness turns out to
be, roughly, equivalent with Kolmogorov randomness: a sequence $x$ is
random if $K(x) \approx l(x)$, i.e. it cannot be effectively
compressed. This theory allows us to speak of the randomness of
single, individual sequences, which is inherently impossible for
probabilistic theories.  Yet, as shown by Martin-L\"of, his notion of
randomness is entirely consistent with probabilistic ideas.
Identifying the randomness of an individual sequence with its
incompressibility opens up a whole new area, which is illustrated by
Example~\ref{ex:posneg}, in which we made distinctions between
different {\em types\/} of random sequences (`positive' and
`negative') that cannot be expressed in, let alone understood from, a
traditional probabilistic perspective.
\paragraph{Philosophy of Statistics/Inductive Inference:
  Epistemological Occam's Razor} There exist two close connections
between algorithmic information theory and inductive inference: one
via the algorithmic sufficient statistic and the MDL Principle; the
other via Solomonoff's induction theory, which there was no space to
discuss here \cite{LiV97}. The former deals with finding structure in
data; the latter is concerned with sequential prediction. Both of
these theories implicitly employ a form of Occam's Razor: when two
hypotheses fit the data equally well, they prefer the simplest one
(with the shortest description).  Both the MDL and the Solomonoff
approach are theoretically quite well-behaved: there exist several
convergence theorems for both approaches. Let us give an example of
such a theorem for the MDL framework: \citeN{BarronC91} and
\citeN{Barron85} show that, if data are distributed according to some
distribution in a contemplated model class (set of candidate
distributions) ${\cal M}$, then two-part MDL will eventually find this
distribution; it will even do so based on a reasonably small sample.
This holds both for practical versions of MDL (with restricted model
classes) as well as for versions based on Kolmogorov complexity, where
${\cal M}$ consists of the huge class of all distributions which can
be arbitrarily well approximated by finite computer programs. Such
theorems provide a justification for MDL. Looking at the proofs, one
finds that the preference for simple models is crucial: the
convergence occurs precisely because complexity of each probabilistic
hyoptheses $P$ is measured by its codelength $L(P)$, under a some
prefix-code that allows one to encode all $P$ under consideration. If
a complexity measure $L'(P)$ is used that does {\em not\/} correspond
to any prefix code, then, as is easy to show, in some situations MDL
will not converge at all, and, no matter how many data are observed,
will keep selecting overly complex, suboptimal hypotheses for the
data. In fact, even if the world is such that data are
generated by a very complex (high $K(P)$) distribution, it is wise to
prefer simple models at small sample sizes \cite{Grunwald07}! This
provides a justification for the use of MDL's version of Occam's razor
in inductive inference. It should be stressed that this is an {\em
  epistemological\/} rather than a {\em (meta-) physical\/} form of
Occam's Razor: it is used as an effective {\em strategy}, which is
something very different from a belief that `the true state of the
world is likely to have a short description'. This issue, as well as
the related question to what extent Occam's Razor can be made
representation-independent, is discussed in great detail in
\cite{Grunwald07}.

A further difference between statistical inference based on
algorithmic information theory and almost all other approaches to
statistics and learning is that
the algorithmic approach focuses on individual data sequences: there
is no need for the (often untenable) assumption of classical
statistics that there is some distribution $P$ according to which the
data are distributed. In the Bayesian approach to statistics,
probability is often interpreted subjectively, as a degree of belief.
Still, in many Bayesian approaches there is an underlying assumption that
there exists `states of the world' which are viewed as probability
distributions. Again, such assumptions need not be made in the present
theories; neither in the form which explicitly uses Kolmogorov
complexity, nor in the restricted practical form. In both cases, the
goal is to find regular patterns in the data, no more. All this is
discussed in detail in \cite{Grunwald07}.

\paragraph{Philosophy of Information}
On the first page of the chapter on Shannon information theory in this
handbook \cite{HarremoesT06}, we read ``information is always {\em
  information about something}.''  This is certainly the case for
Shannon information theory, where a string $x$ is always used to
communicate some state of the world, or of those aspects of the world
that we care about. But if we identify `amount of information in $x$'
with $K(x)$, then it is not so clear anymore what this `information'
is about. $K(x)$, the algorithmic information in $x$ looks at the
information in $x$ itself, independently of anything outside. For
example, if $x$ consists of the first billion bits of the binary
expansion of $\pi$, then its information content  is
the size of the smallest program which prints these bits. This sequence does
not describe any state of the world that is to be communicated.
Therefore, one may argue that it is meaningless to say that `$x$
carries information', let alone to measure its amount. At a workshop
where many of the contributors to this handbook were present, there
was a long discussion about this question, with several participants
insisting that ``algorithmic information misses ``aboutness'' (sic), and
is therefore not really information''. In the end the question whether
algorithmic information should really count as ``information'' is, of
course, a matter of definition. Nevertheless, we would like to argue
that there exist situations where intuitively, the word ``information''
seems exactly the right word to describe what is being measured, while
nevertheless, ``aboutness'' is missing. For example, $K(y|x)$ is
supposed to describe
the amount of ``information'' in $y$ that is not already present in
$x$. Now suppose $y$ is equal to $3 x$, expressed in binary, and $x$
is a random string of length $n$, so that $K(x) \approx K(y) \approx
n$.  Then $K(y|x) = O(1)$ is much smaller than $K(x)$ or $K(y)$. The
way an algorithmic information theorist would phrase this is ``$x$
provides nearly all the {\em information\/} needed to generate $y$.''
To us, this seems an eminently reasonable use of the word information.
Still, this ``information'' does not refer to any outside state of the
world.\footnote{We may of course say that $x$ carries information
  ``about'' $y$. The point, however, is that $y$ is not a state of any
  imagined external world, so here ``about'' does not refer to
  anything external. Thus, one cannot say that $x$ contains
  information about some external state of the world.}

Let us assume then that the terminology ``algorithmic {\em
  information\/} theory'' is justified. What lessons can we draw from
the theory for the philosophy of information?

First, we should emphasize that the amount of `absolute, inherent'
information in a sequence is only well-defined asymptotically and is
in general uncomputable. If we want a nonasymptotic and efficiently
computable measure, we are forced to use a restricted class of
description methods.  Such restrictions naturally lead one to
universal coding and practical MDL. The resulting notion of
information is always defined {\em relative\/} to a class of
description methods and can make no claims to objectivity or
absoluteness.  Interestingly though, unlike Shannon's notion, it is
still meaningful for individual sequences of data, and is not
dependent on any outside {\em probabilistic\/} assumptions: this is an
aspect of the general theory that can be retained in the restricted
forms \cite{Grunwald07}.

Second, the algorithmic theory allows us to formalize the notion of
`meaningful information' in a distinctly novel manner. It leads to a
separation of the meaningful information from the noise in a sequence,
once again without making any probabilistic assumptions. Since
learning can be seen as an attempt to find the meaningful information
in data, this connects the theory to inductive inference.

Third, the theory re-emphasizes the connection between measuring
amounts of information and data compression, which was also the basis
of Shannon's theory. In fact, algorithmic information has close
connections to Shannon information after all, and {\em\/ if\/} the
data $x$ are generated by some probabilistic process $P$, so that the
information in $x$ is actually really `about' something, then
the algorithmic information in $x$ behaves very similarly to the
Shannon entropy of $P$, as explained in Section~\ref{sec:universal}.

\paragraph{Further Reading}
Kolmogorov complexity has many applications which we could not discuss
here. It has implications for aspects of physics such as the second
law of thermodynamics; it provides a novel mathematical proof technique called the {\em incompressibility
  method}, and so on.  These and many other
topics in Kolmogorov complexity are thoroughly discussed and
explained in the standard reference
\cite{LiV97}. Additional (and more recent) material on the relation to
Shannon's theory can be found in
Gr\"unwald and Vit\'anyi 
\citeyear{GrunwaldV03,GrunwaldV05}. Additional material on the structure
function is in \cite{VereshchaginV04,Vitanyi05}; and additional material on
MDL can be found in \cite{Grunwald07}.
\section{Acknowledgments}
Paul Vit\'anyi was supported in part by the
EU project RESQ, IST-2001-37559, the NoE QIPROCONE
+IST-1999-29064 and the ESF QiT Programme.
Both Vit\'anyi and Gr\"unwald were supported in part by the IST Programme of the European
Community, under the PASCAL Network of Excellence,
IST-2002-506778. This publication only reflects the authors' views.
%\bibliography{bib/book,bib/peter,bib/MDL,bib/master,bib/thisvolume}

\end{document}